\documentclass[fleqn,usenatbib,useAMS]{mnras}

\usepackage{newtxtext,newtxmath}
\usepackage[T1]{fontenc}
\usepackage{ae,aecompl}

\usepackage{graphicx}
\usepackage{amsmath}
\usepackage{amssymb}
\usepackage[usenames, dvipsnames]{color}
\usepackage{lscape}
\usepackage{xspace}
\usepackage{ulem}
\usepackage{natbib}
\usepackage{indentfirst}
\usepackage{threeparttable}
\usepackage[capitalize, noabbrev]{cleveref}
\crefformat{section}{\S#2#1#3}
\crefformat{subsection}{\S#2#1#3}
\crefformat{subsubsection}{\S#2#1#3}

\newcommand{\Ha}{\ifmmode {\rm H}\alpha \else H$\alpha$\fi\xspace}
\newcommand{\Hb}{\ifmmode {\rm H}\beta \else H$\beta$\fi\xspace}
\newcommand{\Hba}{\ifmmode {\rm H}\beta^{\prime} \else H$\beta^{\prime}$\fi\xspace}
\newcommand{\Hg}{\ifmmode {\rm H}\gamma \else H$\gamma$\fi\xspace}
\newcommand{\Hd}{\ifmmode {\rm H}\delta \else H$\delta$\fi\xspace}

\newcommand{\Hii}{\ifmmode \rm{H}\,\textsc{ii} \else H\,{\sc ii}\fi}

\newcommand{\Niip}{[N\,{\sc ii}]$\lambda\lambda$6548,6583}
\newcommand{\nii}{\ifmmode [\rm{N}\,\textsc{ii}] \else [N\,{\sc ii}]\fi\xspace}

\newcommand{\oi}{\ifmmode [\rm{O}\,\textsc{i}] \else [O\,{\sc i}]\fi\xspace}
\newcommand{\Oii}{[O\,{\sc ii}]$\lambda$3727}
\newcommand{\neiii}{\ifmmode [\rm{Ne}\,\textsc{iii}] \else [Ne\,{\sc iii}]\fi\xspace}

\newcommand{\heii}{\ifmmode [\rm{He}\,\textsc{ii}] \else [He\,{\sc ii}]\fi\xspace}
\newcommand{\hei}{\ifmmode [\rm{He}\,\textsc{i}] \else [He\,{\sc i}]\fi\xspace}
\newcommand{\oii}{\ifmmode [\rm{O}\,\textsc{ii}] \else [O\,{\sc ii}]\fi\xspace}

\newcommand{\Oiiip}{[O\,{\sc iii}]$\lambda\lambda$4958,5007}
\newcommand{\oiii}{\ifmmode [\rm{O}\,\textsc{iii}] \else [O\,{\sc iii}]\fi\xspace}
\newcommand{\Sii}{[S\,{\sc ii}]$\lambda\lambda$6731,6716}
\newcommand{\sii}{\ifmmode [\rm{S}\,\textsc{ii}] \else [S\,{\sc ii}]\fi\xspace}
\newcommand{\siii}{\ifmmode [\rm{S}\,\textsc{iii}] \else [S\,{\sc iii}]\fi\xspace}
\newcommand{\WHa}{\ifmmode \rm{W}_{{\rm H}\alpha} \else W${}_{{\rm H}\alpha}$\fi\xspace}
\newcommand{\WHacen}{\ifmmode \rm{W}_{{\rm H}\alpha}^{\rm cen} \else W${}_{{\rm H}\alpha}^{\rm cen}$\fi\xspace}
\newcommand{\WHaRe}{\ifmmode \rm{W}_{{\rm H}\alpha}^{\rm Re} \else W${}_{{\rm H}\alpha}^{\rm Re}$\fi\xspace}
\newcommand{\SFRHaSF}{\ifmmode {\rm SFR}_{\rm H\alpha} \else SFR${}_{\rm H\alpha}$\fi\xspace}

\newcommand\starlight{{\sc starlight}}          	
\newcommand{\lpan}{{\it left-panel}}
\newcommand{\rpan}{{\it right-panel}}

\newcommand{\ulpan}{{\it upper-left panel}}

\newcommand{\llpan}{{\it lower-left panel}}
\newcommand{\lrpan}{{\it lower-right panel}}
\newcommand{\Lpan}{{\it Left-panel}}
\newcommand{\Rpan}{{\it Right-panel}}

\newcommand{\ULpan}{{\it Upper-left panel}}
\newcommand{\URpan}{{\it Upper-right panel}}
\newcommand{\LLpan}{{\it Lower-left panel}}
\newcommand{\LRpan}{{\it Lower-right panel}}

\title[AGN hosts on CALIFA]{Galaxies hosting an AGN: a view from the CALIFA survey}

\author[E. A. D. Lacerda et al.]{Eduardo A. D. Lacerda$^{1}$\thanks{E-mail: lacerda@astro.unam.mx}, Sebasti\'an F. S\'anchez$^{1}$, R. Cid Fernandes$^{2}$,
\newauthor
Carlos L\'opez-Cob\'a$^{1}$, Carlos Espinosa-Ponce$^{1}$, L. Galbany$^{3}$,
\\
$^{1}$Instituto de Astronom\'ia, Universidad Nacional Aut\'onoma de M\'exico, A.P. 70-264, C.P. 04510, CDMX, M\'exico.\\
$^{2}$Departamento de F\'isica-CFM, Universidade Federal de Santa Catarina, C.P. 476, 88040-900, Florian\'opolis, SC, Brazil.\\
$^{3}$Departamento de F\'isica Te\'orica y del Cosmos, Universidad de Granada, E-18071 Granada, Spain.
}

\date{Accepted 2019 December 29. Received 2019 December 19; in original form 2019 October 18}

\pubyear{2019}

\begin{document}
\label{firstpage}
\pagerange{\pageref{firstpage}--\pageref{lastpage}}
\maketitle

\begin{abstract}
We study the presence of optically-selected Active Galactic Nuclei (AGNs) within a sample of 867 galaxies extracted from the extended {\it Calar-Alto Legacy Integral Field spectroscopy Area} (eCALIFA) spanning all morphological classes. We identify 10 Type-I and 24 Type-II AGNs, amounting to $\sim4$ per cent of our sample, similar to the fraction reported by previous explorations in the same redshift range. We compare the integrated properties of the ionized and molecular gas,
and stellar population
of AGN hosts and their non-active counterparts, combining them with morphological information. The AGN hosts are found in transitory parts (i.e. green-valley) in almost all analysed properties which present bimodal distributions (i.e. a region where reside star-forming galaxies and another with quiescent/retired ones). Regarding morphology, we find AGN hosts among the most massive galaxies, with enhanced central stellar-mass surface density in comparison to the average population at each morphological type. Moreover, their distribution peaks at the Sab-Sb classes and none are found among very late-type galaxies (> Scd). Finally, we inspect how the AGN could act in their hosts regarding the quenching of star-formation.
The main role of the AGN in the quenching process appears to be the removal (or heating)
of molecular gas, rather than an additional suppression of the already observed decrease
of the star-formation efficiency from late-to-early type galaxies.

\end{abstract}

\begin{keywords}
galaxies: nuclei -- galaxies: evolution -- galaxies: star formation
\end{keywords}


\section{Introduction}
\label{sec:intro}

It is likely that all local galaxies (or at least those with a bulge) harbour a supermassive black hole (SMBH) at the centre that, when accretes surrounding gas, ignites an Active Galactic Nucleus \citep[AGN;][]{Kormendy.and.Douglas.1995}. The energy released in this natural process has been interpreted as a plausible cause of the halting of star-formation (SF) in galaxies \citep[e.g.][]{Silk.and.Reese.1998} via removal of the interstellar cold gas \citep[e.g.][]{Fabian.2012,Trussler.etal.2018} or heating it to a temperature that makes it unfeasible to cool down in a considerable time-scale \citep[e.g.][]{Bower.etal.2006}. Conversely, in some cases, it could promote SF by creating disturbances that help the fragmentation of cold gas clouds \citep[e.g.][]{Silk.2005}. Apart from being negative or positive, the AGN feedback should couple the evolution of the SMBH with their host galaxy \citep[e.g.][]{Silk.and.Reese.1998}, affecting the amount of gas in the interstellar medium and the properties of the underlying stellar population.

The bimodality of galaxies is under the spotlight of evolution at the present epoch. The luminosity coming from red galaxies at least double up since z$\sim$1 \citep[e.g.][]{Bell.etal.2004, Faber.etal.2007} as a physical manifestation of the growth of the red sequence population \citep[e.g.][]{SFS.etal.2019a}. This fundamental bifold behaviour tells us that the present-day red galaxies ceased their SF at some time in the past. Their differences could be perceived in the color-magnitude diagram \citep[CMD; e.g.][]{Strateva.etal.2001, Baldry.etal.2004}, stellar mass \citep[e.g.][]{Gallazzi.etal.2005, Gallazzi.etal.2008}, star-formation rate \citep[SFR; e.g.][]{Brinchmann.etal.2004, Daddi.etal.2007, Renzini.and.Peng.2015}, both global and local dominance over the ionization budget \citep[e.g.][]{Stasinska.etal.2008, Sarzi.etal.2010, CF.etal.2010, CF.etal.2011, Lacerda.etal.2018}, among other properties. The role/relevance of AGNs in this separation is not completely understood. However, when hosting such phenomena, galaxies experience different behaviors reflected on their stellar and gas properties, probably accounting for such observed segregation \citep[e.g.][hereafter S18]{Kauffmann.etal.2007, SFS.etal.2018a}. Furthermore, AGN hosts populate the green-valley of the CMD (e.g. \citealt{Kauffmann.etal.2003c, SFS.etal.2004a, Martin.etal.2007, Salim.etal.2007}; S18) and SFR-M$_\star$ diagram (e.g.  \citealt{Schawinski.etal.2014}; S18), reinforcing its fleeting behavior.

In addition, there is a tight correlation between the central BH mass and the velocity dispersion of the stellar population in galaxies \citep[M${}_{\rm BH}-\sigma_\star$ relation;][]{Ferrarese.and.Merritt.2000, Gebhardt.etal.2000}. This points in the direction of a coevolution between the SMBH and its host galaxy. This idea seems to be right at certain level, as the coevolution exists only when a bulge is present \citep[e.g.][and references therein]{Greene.etal.2008, Greene.etal.2010, Kormendy.and.Ho.2013}. Moreover, the growth of the bulge could explain the dominance of inside-out over outside-in quenching \citep[e.g.][]{GonzalezDelgado.etal.2016, Ellison.etal.2018, Lin.etal.2019, Bluck.etal.2019}. Thus, there seem to be a connection between AGN activity, bulge and SMBH growth and SF quenching.

So far it is considered that the AGN negative feedback described before (i.e., the injection of energy from the nuclear source onto the host galaxy) is the main driver of the suggested connection. Negative feedback implies the removal (mechanical feedback) or heat (thermal feedback) of the cold gas, preventing the formation of new stars. Nowadays, the presence of this negative feedback is almost mandatory in numerical, theoretical and semi-analytic models in order to explain the lack of massive galaxies in the high-mass end of the mass function \citep[e.g.][]{Kauffmann.and.Haehnelt.2000, DiMatteo.etal.2005, Croton.etal.2006, Somerville.etal.2008, Schaye.etal.2015, Sijacki.etal.2015, Lacey.etal.2016}. The most accepted evolutionary scenario between star-forming galaxies (SFGs) and retired galaxies (RGs) implies the ignition of an AGN that effectively quench the SF, and that ignition is triggered by a process that modifies the morphology too \citep[i.e., a major merger;][]{Hopkins.etal.2010}. However, that simple picture lacks strong observational support, since host galaxies of AGNs are not preferentially found in highly disturbed or clear merging systems \citep[e.g.][]{SFS.etal.2014}.

For all these reasons, it is important to compare the properties of AGN hosts with those of non-active galaxies to understand the nature of such connection. Here we explore if AGN hosts are in a transitory phase between SFGs and RGs. We will try to understand the role of AGN activity in causing the SF quenching, or if the quenching is produced by a different process that co-evolves with the ignition of nuclear activity \citep[e.g., bulge growth that stabilizes the disk, hampering the fragmentation of molecular clouds,][]{Martig.etal.2009}. For doing so, we select a sample of AGN hosts in the nearby Universe ($z<0.1$) based on a dataset of Integral Field Spetroscopy (IFS) provided by the {\it Calar-Alto Legacy Integral Field spectroscopy Area} survey \citep[CALIFA,][]{SFS.etal.2012}. This provides us with a census of the local strong optical AGNs. We characterize the main properties of their host galaxies and the differences between them and their non-active counterparts. Finally, we explore the presence of AGN and its relation with two mechanisms explored by \cite{Bitsakis.etal.2019}: (i) the lack of cold gas and (ii) the decrease of the efficiency to form new stars. In addition, we include the morphology of the host galaxy in the analysis, searching for a morphological connection with the aforementioned mechanisms.


This paper is organized as follows. In \cref{sec:data} we summarize the general properties of the employed sample of galaxies. The stellar population synthesis, the fit of the emission lines and the AGN selection scheme are unfolded at \cref{sec:analysis}. The main properties of AGN hosts and the comparison with their non-active counterparts appear in \cref{sec:results}. \cref{sec:disc} brings a comparison with other recent IFS AGN studies and the discussion about what role the AGN plays in the evolution of their host galaxies and we finalize with our conclusions at \cref{sec:summary}. Along this article we assume the standard $\Lambda$ Cold Dark Matter cosmology with the parameters: H$_0$=71 km/s/Mpc, $\Omega_M$=0.27, $\Omega_\Lambda$=0.73.

\section{Data}
\label{sec:data}

The dataset adopted along this study comprises all the observations with good quality obtained by the CALIFA survey in the low resolution mode (V500), together with all the CALIFA extended surveys \citep{CALIFADR3, Galbany.etal.2018}, including data from the PMAS/Ppak Integral-field Supernova hosts COmpilation (PISCO; \citealp{Galbany.etal.2018}). This sample (eCALIFA hereafter), comprises 867 galaxies selected following the criteria of the original mother sample \citep[MS;][]{Walcher.etal.2014}, but relaxing some of them. In general, all of galaxies were diameter selected, with most of their optical extension fitting within the field-of-view (FoV) of the instrument, but they could be either fainter or brighter than the limits adopted for the MS, or located at slightly larger redshifts, slightly out of the boundaries of the original selection. The main difference with the original mother sample is that the selection is not restricted to the SDSS foot-print \citep{York.etal.2000}. The final sample comprises galaxies of any morphological type, covering the mass range between $10^{7.6} - 10^{11.9}$\,M$_\odot$ (mean $10^{10.5}$\,M$_\odot$), and a redshift range between $0.001 < z < 0.08$ (with 93 per cent of them with $z<0.035$, i.e., a similar redshift foot-print of the original CALIFA sample). In general they comprise a representative sample of the galaxies in the nearby Universe.

Observations are made using the V500 setup of the Integral Field Unity (IFU) instrument (PMAS/PPAK, \citealt{Roth.etal.2005, Kelz.etal.2006}) at the Calar Alto 3.5\,m telescope. This configuration results in a spectral range covering 3745-7500\,\AA\ 
with a nominal resolution of $\lambda/\Delta\lambda \sim 850$ at 5000 \AA\ (FWHM $\sim 6$\,\AA). A three point dithering scheme was performed in order to increase the spatial resolution and cover 100 per cent of the field-of-view ($74'' \times 64''$). Data were reduced using version 2.2 of the CALIFA reduction pipeline. The processes involved in the reduced pipeline are described in the third data-release of the CALIFA survey \citep[DR3,][]{CALIFADR3}. The final product of the reduction is a datacube with two dimensions corresponding to the spatial coordinates (right-ascension and declination), and a spatial sampling of 1$\arcsec$/spaxel, and the third dimension corresponding to the spectral range. The galaxies from the sample are covered well enough to obtain integrated and spatial resolved properties probing up to 3 (in some cases 4) effective radii. At the distances of our sources, the estimated spatial PSF FWHM ($\sim 2.5''$) is around 0.8 kpc in average.

Morphological information for all the objects in the original CALIFA mother sample was retrieved from \citet{Walcher.etal.2014}, comprising 635 of the current sample. For the remaining 232 objects in the eCALIFA sample we adopted the same procedure described in that article to derive the morphological classification by eye. In summary we inspected the true-color SDSS \citep[DR7,][]{Abazajian.etal.2009} images, and when not available we used similar images extracted from the datacubes (at lower spatial resolution), together with emission line images (to detect possible traces of spiral arm structures). As indicated before, the final sample spans over all Hubble types, comprising 163 ellipticals (E0-E7), 105 lenticulars (S0+S0a), 590 spirals (Sa-Sm) and 9 irregular (I) galaxies.

\section{Analysis}
\label{sec:analysis}

\subsection{Stellar population synthesis and Emission-line fitting}
\label{sec:analysis:syntheml}

The datacubes were analysed using the {\sc Pipe3D} pipeline \citep{SFS.etal.2016b}, in order to extract information of the properties of the stellar population and the emission lines. Its current implementation adopts the GSD156 library of simple stellar populations, SSP \citep{CF.etal.2013}. This SSP-library comprises 156 templates for 39 stellar ages from 0.001 to 14.1\,Gyr and four metallicities ($Z/Z_\odot$ = 0.2, 0.4, 1 and 1.5), that adopts the Salpeter initial mass function \citep[IMF; ][]{Salpeter.1955}. This library has been widely used by various previous publications \citep[e.g.][]{Perez.etal.2013, GonzalezDelgado.etal.2014b, SanchezMenguiano.etal.2016, SanchezMenguiano.etal.2018, Lin.etal.2019, IbarraMedel.etal.2019}. Details of the fitting procedure, dust attenuation curve, uncertainties of the processing of the stellar populations and emission lines, and the dataproducts derived by {\sc Pipe3D} are included in \citet{SFS.etal.2016}, \citet{SFS.etal.2016b}. We include here a brief summary of the main processes included in the pipeline.

\begin{figure*}
    \includegraphics[width=1.0\textwidth]{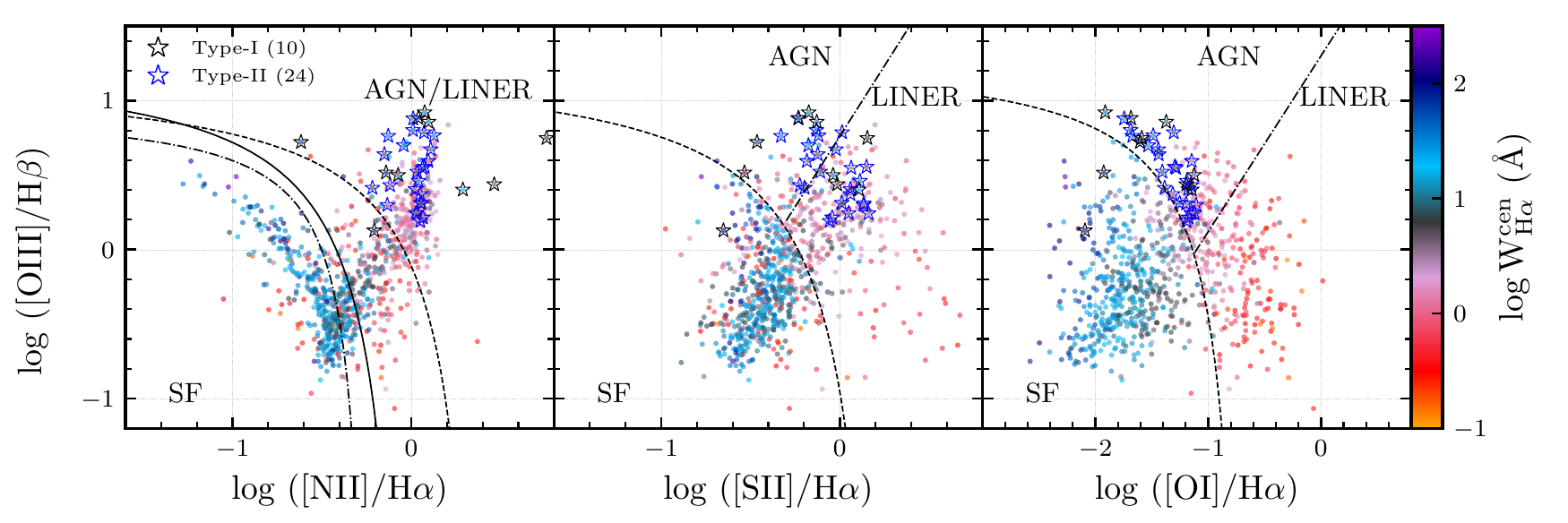}
    \caption{Diagnostic diagrams using optical emission lines ratios (\nii/\Ha, \sii/\Ha and \oi/\Ha versus \oiii/\Hb) from the central region ($3''\times3''$) of the galaxies in the eCALIFA sample. The dividing dashed curves inside all diagrams come from \citet[K01]{Kewley.etal.2001a}. The dot-dashed curve on the first diagram (\nii/\Ha), from \citet[S06]{Stasinska.etal.2006a}. The solid line on the same diagram, from \citet[K03]{Kauffmann.etal.2003c}. Finally, the dot-dashed from two diagrams to the right (\sii/\Ha and \oi/\Ha), dividing AGN and LINER regions, from \citet[K06]{Kewley.etal.2006}. The colors show the central EW of \Ha, \WHacen. The colored stars mark our {\it bona fide} AGN candidates (Type-I as black open stars and Type-II as blue open stars). The color figure can be viewed online.}
    \label{fig:BPTVO}
\end{figure*}

First, the stellar population is analysed. To do so the datacube is spatially binned in the V-band to increase the signal-to-noise (S/N) above the limit at which the derived parameters are reliable \citep[S/N$\sim$50, based on][]{SFS.etal.2016}. However, in order to preserve the original shape of the galaxies and minimize the mixing of different physical regions, the adopted binning impose a maximum difference in the surface-brightness between adjacent spaxels \citep{SFS.etal.2016b}. Then a stellar population fit is applied to the co-added spectra within each spatial bin using the {\sc Fit3D} code. A spaxel-wise stellar-population model is estimated then by re-scaling the best fitted model (in each bin) to the continuum flux intensity in the corresponding spaxel \citep{CF.etal.2013, SFS.etal.2016}. The model is then used to derive the spaxel-wise stellar mass density and co-adding through the entire FoV, the total stellar mass of the galaxy.

A gas-pure datacube is created by subtracting the spaxel-wise stellar-population model from the original cube. This cube is used to derive the spatial resolved properties of the ionized gas emission lines. A moment analysis is performed for a set of 52 emission lines \citep{SFS.etal.2016b, SFS.etal.2018a}, deriving the flux intensity, velocity, velocity dispersion and equivalent width (EW) for each individual spectrum. The final dataproducts comprise a set of maps for each galaxy, one for each of the estimated parameter (with its corresponding error). Along this article we use only the fluxes and EWs for the strongest emission lines:
\Ha, \Hb, \Oii, \Oiiip, \Niip, \Sii. Each of these lines were corrected for dust attenuation in a post-processing of the data. The dust attenuation was estimated spaxel-wise, using the \Ha/\Hb\ ratio, assuming a \citet{Cardelli.etal.1989} extinction law with R$_{\rm V}$=3.1 and a standard value of the intrinsic ratio of 2.86 \citep{Osterbrock.1989}.

Along this paper we derive the SFR from the \Ha luminosity calculated spaxel-by-spaxel with detected ionized gas for each galaxy, using the dust-corrected \Ha luminosity and then converted to SFR following the \citet{Kennicutt.1998} calibration. This procedure considers spaxels irrespective of their ionization source, thus a part of the observed \Ha is not coming from SF. This \Ha emission contamination is discussed by \citet{CanoDiaz.etal.2016} and \citet{SFS.etal.2017} and evaluated by \citet{Lacerda.etal.2018}. The SFR value is minorly affected by this contamination in galaxies dominated by SF, as previously noted by \citet{CatalanTorrecilla.etal.2017}. Conversely, galaxies with ionization budget dominated by other processes, such as AGN, jets, shocks and post-AGB (Asymptotic Giant Branch) stars  would be affected and their derived SFR should be consider as an upper limit. In order to mitigate part of this caveat, in our version of the \Ha-based SFR, the \Ha flux is decontaminated for post-AGB ({\it hot low-mass evolved stars}; \citealt{FloresFajardo.etal.2011}) emission by assuming that these sources add 1 \AA\ to the EW of \Ha, \WHa \citep{Binette.etal.1994, Stasinska.etal.2008, CF.etal.2011}.

The cold gas is a key property for the understanding of the evolution of galaxies since it is the fuel of the SF \citep[e.g.][]{Kennicutt.and.Evans.2012, Krumholz.etal.2012}. Recent studies have evaluated how and why the SF stops using different data, methods and assumptions \citep[e.g.][]{Cappellari.etal.2013, GonzalezDelgado.etal.2014a, Casado.etal.2015, Saintonge.etal.2016}. The relation between the SFR and the density of the interestellar gas proposed by \citet{Schmidt.1959}, later known as the Schmidt-Kennicutt law \citep[SK-law][]{Kennicutt.1998}, is maintained at kpc-scales only for the molecular gas \citep[e.g.][]{Kennicutt.etal.2007, Bigiel.etal.2008, Leroy.etal.2013}. Important efforts have been made in order to gather direct measurements of molecular gas in galaxies present in the CALIFA sample by the EDGE-CALIFA collaboration \citep[e.g.][]{Bolatto.etal.2017}, but so far they have collected CO measurements for only 126 galaxies. We have thus estimated the molecular gas content indirectly by means of a gas-to-dust conversion with a correction factor that depends of the oxygen abundance as derived by S18:

\begin{equation}
    \Sigma_{\rm gas} [{\rm M}_{\odot}/{\rm pc^2}] = 15 \left(\frac{A_V}{mag}\right) + \left[\log({\rm O}/{\rm H}) - 2.67\right].
\end{equation}

Dust extinction along the line-of-sight correlates with the molecular hydrogen column density \citep[e.g.][]{Dickman.1978, Bohlin.etal.1978, Rachford.etal.2009, Brinchmann.etal.2013} allowing us to estimate the gas content in galaxies. A detailed study of the spatially-resolved distribution of the total cold gas (atomic+molecular) from the optical dust attenuation calibration is presented by \citet{BarreraBallesteros.etal.2019arXiv}. Following those results, the molecular gas mass surface density, $\Sigma_{\rm gas}$, is estimated by the local (i.e. spaxel-by-spaxel) gas-to-dust conversion and integrated for the entire FoV of each galaxy. On average, these estimated surface density values, when compared with CO-based measurements, present a dispersion of $\sim0.3$\,dex (e.g. \citealt{Galbany.etal.2017}), therefore our integrated gas masses values should be considered as a first order approximation to the real ones.

Further global parameters derived from the emission lines or directly from the multi-SSP analysis of the stellar population synthesis, like both characteristic oxygen abundance and stellar metallicity (thus, at the effective radius) are derived following the same procedure described in several previous articles \citep[e.g.][]{SFS.etal.2017,SFS.etal.2018a,SFS.etal.2019b}. We refer the reader those articles for the details on their particular derivation, to avoid unnecessary repetitions.

\subsection{AGN candidates}
\label{sec:analysis:candidates}

The ionized gas located in the central region of galaxies carries the spectroscopic signatures of the presence or not of an active nucleus. The analysis of diagnostic diagrams based on flux ratios between emission lines observed in the optical range have helped us through years to this goal (e.g. \citealt*{BPT.1981}; \citealt*{Veilleux.Osterbrock.1987}; \citealt{Veilleux.etal.1995}; \citealt{Kewley.etal.2001a}; \citealt{Kauffmann.etal.2003c}). Ratios between collisionally excited to recombination lines are sensitive to electronic temperature so that the hardness of the ionization field can be measured combining them (e.g. \nii/\Ha vs \oiii/\Hb, \oi/\Ha vs. \oiii/\Hb).

In pursuance of AGN host candidates among our sample, we plot in \cref{fig:BPTVO} the distribution of \nii/\Ha, \sii/\Ha, \oi/\Ha vs \oiii/\Hb emission-line ratios (here we call it BPTVO, after \citealt*{BPT.1981} and \citealt*{Veilleux.Osterbrock.1987}) integrated over the central $3'' \times 3''$ region of 849, 837 and 834 galaxies respectively. They comprise all galaxies with the considered emission lines detected at the central aperture above a 3$\sigma$ detection limit. Combining them we obtain 819 objects with all needed lines measured. For the remaining 48 galaxies, at least one of the considered emission lines is not detected in this particular aperture. The numbers are different between diagrams due to the natural differences in the line intensity of each emission-line, and therefore the ability to detect them. From the initial sample of 867 galaxies, we do not detect \Ha (the strongest line, in general) in the central region of only 10 of them.

\begin{figure*}
    \includegraphics[width=\columnwidth]{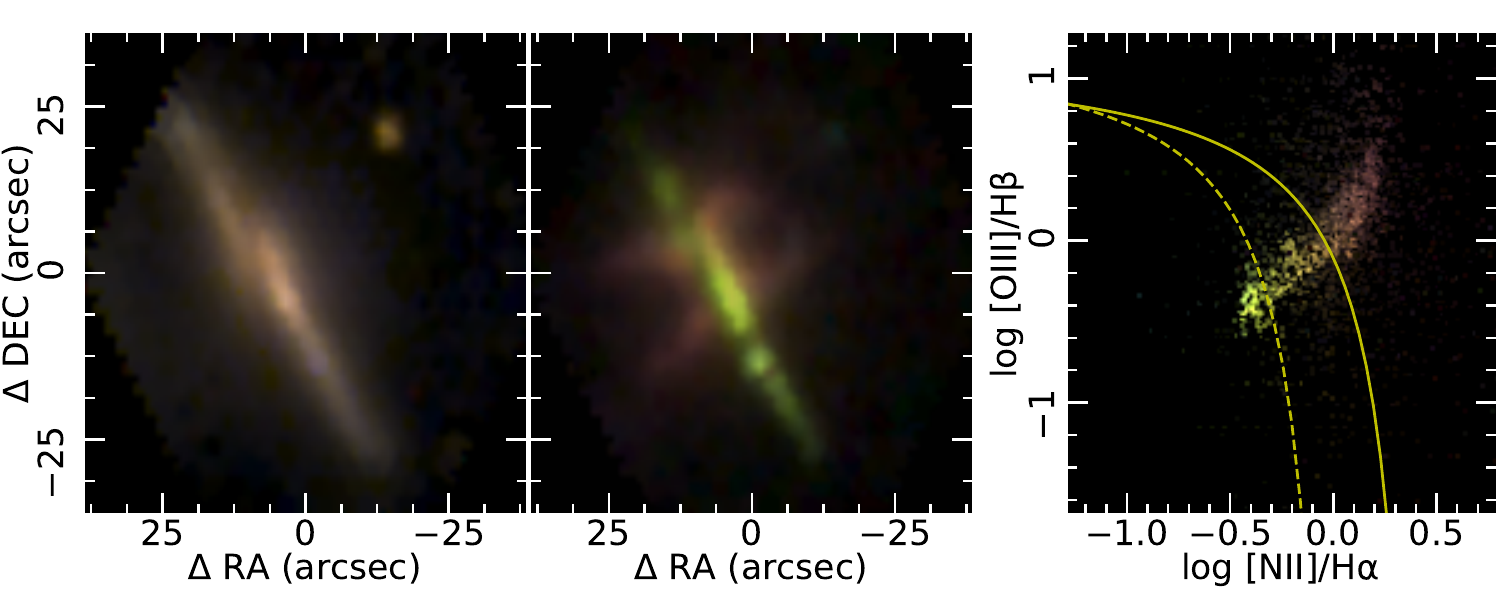}
    \includegraphics[width=\columnwidth]{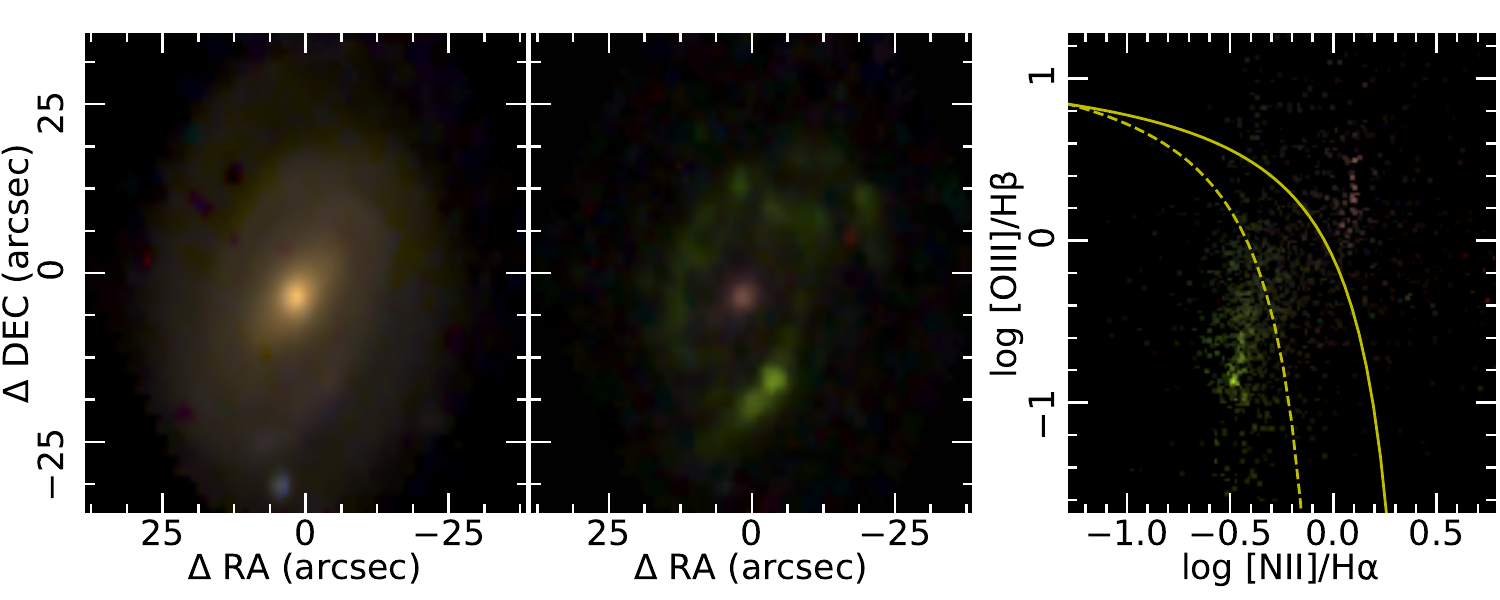}
    \includegraphics[width=\columnwidth]{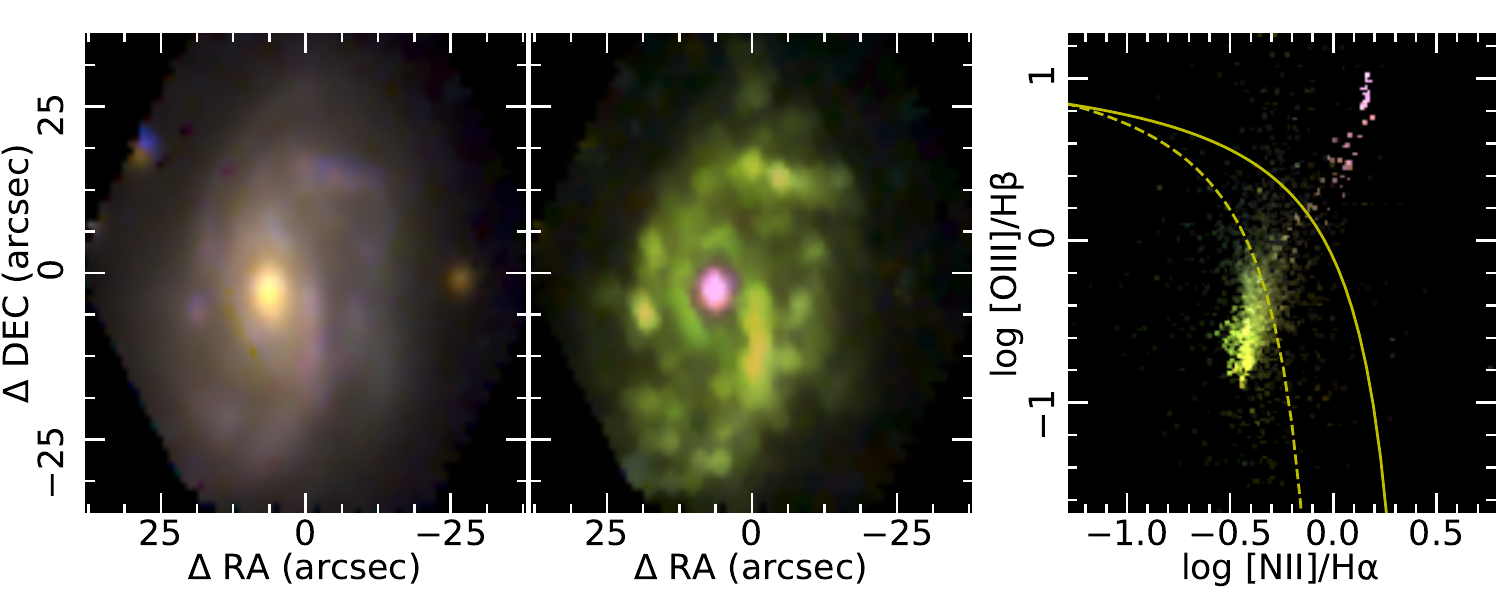}
    \includegraphics[width=\columnwidth]{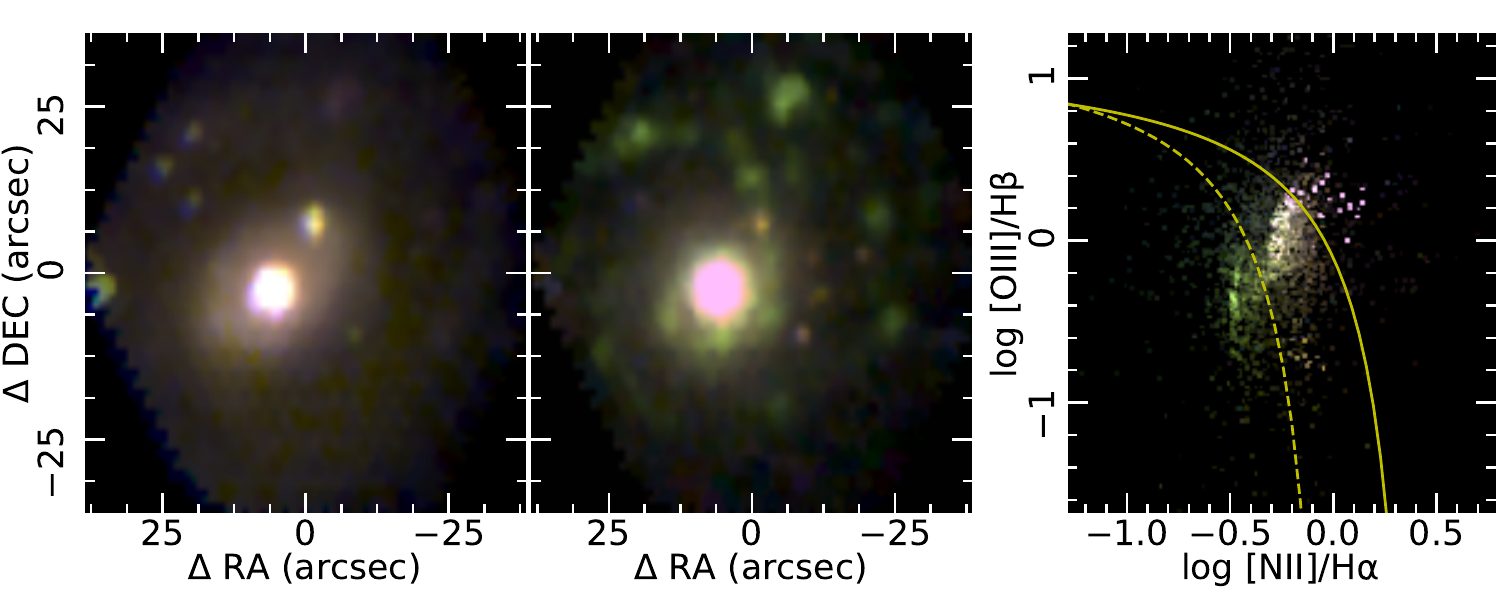}
    \caption{Four examples of galaxies in our sample. Each galaxy is shown in a triad of diagrams: two maps and a BPT plane. The leftmost panel shows the combined optical RGB image of the galaxy continuum; the central is the main emission lines RGB map (\nii\ as red, \Ha as green and \oiii as blue) and the rightmost is the BPT plane, which carries the spatial information shown in the central map. Thus, each point in the BPT diagram corresponds to a pixel in the leftmost and middle panels, with the colors of the middle one. The first galaxy is NGC\,6286, a non-active galaxy which suffers from a SF-driven outflow (\WHacen = 28\.\AA) and is plagued by shock ionization, which produces line ratios very similar to those from AGN ionization. The galaxy IC\,4566 appears in upper-right triad of panels. This object is classified by the BPT diagram as an AGN host candidate. However, its \WHacen is equal to 1.78\,\AA, i.e., its central ionization is more likely due to old stars \citep{Singh.etal.2013, Lacerda.etal.2018}, the reason which is excluded from our final AGN sample. Both galaxies in the second row are AGN hosts. The left one is IC\,2247, a Type-II AGN host and the right one, NGC\,7469, a face on Type-I AGN host with a clear broad component in emission lines of the Balmer series in the central spectra. This latter galaxy is one of the two black stars in our final sample which are in the SF-loci of all diagnostic diagrams in the BPTVO. The color figure can be viewed online.}
    \label{fig:examples}
\end{figure*}

The lines in each diagram included in \cref{fig:BPTVO} were designed to indicate the main ionization source of different {\it loci} over the analysed plane. The \citet[K01]{Kewley.etal.2001a} and the \citet[K03]{Kauffmann.etal.2003c} lines are the most known. They are typically used to distinguish between pure star-forming galaxies (below the K03 line) and pure AGN/LINER (Low-ionization Nuclear Emission-line Region) hosts (above the K01 line). However, with the spread usage comes a lot of misleading interpretation of its purposes and real foundations. The K01 line was designed to select galaxies that harbour an ionization different than SF with certainty. Its derivation was based on photoionization models covering a wide range of parameters and stellar population synthesis models. It traces the maximum values of the involved line ratios that can be generated by ionization due to young stars (thus, associated with SF). Objects above it have to harbour an additional ionization source (different from what they considered SF), however its nature is not well determined. In general AGNs are located above this line \citep[e.g.][]{Osterbrock.1989}, therefore central ionization found in this location of the diagram is in general associated with the presence of an AGN. Nonetheless, other sources of ionization may populate these regions too. The most frequent ones are: (i) diffuse ionization ubiquitous associated with old-stellar populations \citep[e.g.,  post-AGBs;][]{Binette.etal.1994, Singh.etal.2013, Gomes.etal.2016b} as shown in the central region of the galaxy at the upper-right panel in \cref{fig:examples}, IC\,4566; and (ii) shock ionization, associated with galactic outflows at the resolution of our data \citep[e.g.][]{BlandHawthorn.etal.2007}. This latter example could be seen in galaxy at the upper-left panel in \cref{fig:examples}. NGC\,6286 is an interacting galaxy hosting outflows identified by the study of \citet{LopezCoba.etal.2019} and it has the central EW of \Ha, \WHacen, equals to 28\,\AA, but a soft central spectrum compatible with ionization by SF. However, this galaxy is full of regions dominated by the hard ionization compatible with AGN/shocks in the outskirts.

In addition, stellar population synthesis including the near-infrared (NIR) modeling shows that Type-I AGN hosts have a contribution of young stellar components up to 90 per cent in the central spectra \citep[e.g.][]{Riffel.etal.2009}. This effect could move a source to the SF region of the diagnostic diagrams applyied here. On the other hand, distinguishing between the presence of an AGN from post-AGB ionization it is possible by introducing a cut in the \WHa \citep[e.g.][at least for the strong AGNs]{CF.etal.2011, Lacerda.etal.2018}. On the other hand, shock ionization induced by a galactic outflow can be identified by examining the morphology of the ionized gas distribution towards the minor axis, where its typical distribution adopts conical/biconical shapes \citep{LopezCoba.etal.2017a, LopezCoba.etal.2019}. Optical counterparts of radio-jets and supernova remnants may be also located in this region of the diagram, however they are far less frequent.  

\subsubsection{The selection of Type-II AGN candidates}
\label{sec:analysis:candidates:tIIcriteria}

The issue of identifying the ionizing source in galaxies or regions of galaxies based just on pure line ratios has been discussed extensively in previous studies \citep[e.g.][]{SFS.etal.2012, CanoDiaz.etal.2016, LopezCoba.etal.2019}.
Based on the prescriptions adopted in these articles and adapting the recipe used by S18, we select our sample of {\it bona-fide} AGN candidates (black and blue open stars in \cref{fig:BPTVO}). This involves a combination of the {\it loci} within the BPTVO diagrams together with a limit in \WHacen. If instead of that criteria, we had only adopted a demarcation line (even the more conservative one, i.e., K01) as our boundary between AGNs and SF ionization, we would have selected 181/259/312 AGN hosts respectively (for each diagram). Conversely, we can combine the BPTVO diagram together with one of the other two, selecting objects simultaneously above the two involved demarcation lines. With this selection we would have found 169 AGNs using the first diagram plus the second one (BPT + \sii/\Ha) and 131 objects using the first plus the third one (BPT + \oi/\Ha). A lot of these objects are certainly not AGN hosts, based on the results by \citet{Stasinska.etal.2008} and \citet{CF.etal.2010}, since they have low \WHacen values. In order to remove objects with the central ionization source most probably coming from post-AGBs, we select only those galaxies that have an \WHacen above 3\,\AA. An example of this cut effect could be seen in \cref{fig:examples} where the galaxy IC\,4566 (upper-right triad of panels) and NGC\,1667 (lower-left triad of panels) are very similar regarding RGB images and BPT planes and both could be classified by AGN hosts if only the BPT plane of the central spectrum was considered. However, IC\,4566 has the central ionization dominated by HOLMES, i.e., a hard ionizing field with low-\WHa, $\sim$1.8\,\AA. On the other hand, NGC\,1667 has \WHacen $\sim$11.4\,\AA, showing a signature of a strong AGN. After this cut we obtain, in the same order described before, 69, 90 and 49 galaxies, using only one of the diagrams, and 60 and 34 using the BPT plus \sii/\Ha or \oi/\Ha. Therefore, following S18, to choose our {\it bona-fide} Type-II AGN candidates, we select our candidates in the most restrictive criteria: (i) they have to lie above the K01 line in all three diagrams at the same time; and (ii) have \WHacen larger than 3\,\AA\ \citep[e.g.][]{CF.etal.2011}. S18 adopted a slightly less restrictive criterion for the \WHacen (1.5\AA), due to the lower spatial resolution of the MaNGA sample, and the involved dilution \citep[e.g.][]{Mast.etal.2014} and possible contamination by other sources of ionization for the same central aperture selected here. From the 63 initial candidates above the K01 line in the BPT and with \WHacen > 3\,\AA, 60 of them are also above the K01 line in the \sii/\Ha diagram. Finally, of those 60 candidates, 32 lay above the K01 line in the \oi/\Ha diagram. These are the initial sample of AGN host candidates. Raising the \WHacen\ limit to 6\,\AA, we loose 12 Type-II candidates. This last value was attributed as the borderline between weak and strong AGN (wAGN and sAGN) {\it loci} in the WHAN diagram \citep{CF.etal.2011}. Thus, of the 32 candidates, 20 are sAGNs and 12 would be wAGNs.

Note that we do not use the \citet[K06]{Kewley.etal.2006} line, usually adopted to segregate between AGN and LINERs, since this segregation occurs naturally by the limit adopted in the \WHacen. Of the 32 candidates, 18 are below the AGN/LINER division in the \sii/\Ha diagram, but none are below this demarcation line in the \oi/\Ha diagram, showing that this latter is a better proxy to the strength of the ionization field (e.g. \citealt{Schawinski.etal.2010}; S18). As noted by these two previous studies, we see that our AGN selection disagrees with the Seyfert/LINER separation line in the BPT diagram \citep[K06 line updated by][]{CF.etal.2011}, with 41 per cent of the selected candidates to host an AGN located below it. For the K06 line in the \sii/\Ha diagram we found 53 per cent below it. Although this foregoing procedure could bias our sample towards hosts of strong AGNs, as we need to measure various emission lines and make a restriction to the \WHacen, such method helps us to select more reliable candidates and not all possible contenders. Our final sample of {\it bona-fide} Type-II AGN hosts is composed by 24 candidates (blue open stars on \cref{fig:BPTVO}). The 8 remaining candidates (from the 32 that passed all criteria to select an AGN host in this study) are classified as Type-I AGN, as described in the next Section.  

\subsubsection{Type-I AGN candidates}
\label{sec:analysis:candidates:tIcriteria}

\begin{table*}
    \centering
    \caption{Final sample of AGN hosts$^1$.}
    \begin{threeparttable}
        \begin{tabular}{c|c|c|c|c|c}
            \hline
            Galaxy Name & AGN Type & RA & DEC & redshift & Morphology \\
            \hline
            \hline
            UGC\,01859 & 1 & 02h24m44s & +42d37m24s & 0.019872 & E4 \\
            UGC\,03973 & 1 & 07h42m33s & +49d48m34s & 0.022048 & Sbc \\
            UGC\,03995 & 1 & 07h44m09s & +29d14m51s & 0.015735 & Sb \\
            CGCG\,064-017 & 1 & 09h59m15s & +12d59m02s & 0.034273 & Scd \\
            NGC\,5929 & 1 & 15h26m06s & +41d40m14s & 0.008385 & Sb \\
            NGC\,6251 & 1 & 16h32m32s & +82d32m16s & 0.024440 & E5 \\
            NGC\,6264 & 1 & 16h57m16s & +27d50m59s & 0.033883 & Sab \\
            UGC\,11680\,NED01 & 1 & 21h07m41s & +03d52m18s & 0.025800 & Sb \\
            NGC\,7469 & 1 & 23h03m16s & +08d52m26s & 0.016169 & Sc \\
            UGC\,12505 & 1 & 23h19m25s & +05d54m22s & 0.020278 & Sc \\
            \hline
            MCG\,-02-02-030 & 2 & 00h30m07s & -11d06m49s & 0.011787 & Sb \\
            UGC\,00934 & 2 & 01h23m27s & +30d46m42s & 0.034871 & Sb \\
            UGC\,00987 & 2 & 01h25m31s & +32d08m12s & 0.015301 & Sa \\
            NGC\,0833 & 2 & 02h09m21s & -10d07m59s & 0.012702 & Sa \\
            PGC\,009572 & 2 & 02h30m51s & +22d28m13s & 0.030990 & E4 \\
            NGC\,1093 & 2 & 02h48m16s & +34d25m13s & 0.017428 & Sbc \\
            NGC\,1667 & 2 & 04h48m37s & -06d19m12s & 0.015049 & Sbc \\
            UGC\,03789 & 2 & 07h19m31s & +59d21m18s & 0.010692 & Sb \\
            NGC\,2410 & 2 & 07h35m02s & +32d49m20s & 0.015591 & Sb \\
            IC\,2247 & 2 & 08h16m00s & +23d11m59s & 0.014242 & Sab \\
            NGC\,2554 & 2 & 08h17m54s & +23d28m20s & 0.013507 & S0a \\
            NGC\,2639 & 2 & 08h43m38s & +50d12m20s & 0.010502 & Sa \\
            IC\,0540 & 2 & 09h30m10s & +07d54m10s & 0.006701 & Sab \\
            NGC\,3160 & 2 & 10h13m55s & +38d50m35s & 0.022733 & Sab \\
            UGC\,06719 & 2 & 11h44m47s & +20d07m24s & 0.021831 & Sb \\
            NGC\,3861 & 2 & 11h45m04s & +19d58m25s & 0.016874 & Sb \\
            NGC\,5216 & 2 & 13h32m07s & +62d42m02s & 0.009742 & E0 \\
            NGC\,5443 & 2 & 14h02m12s & +55d48m50s & 0.005982 & Sab \\
            NGC\,5533 & 2 & 14h16m08s & +35d20m38s & 0.012793 & Sab \\
            NGC\,5675 & 2 & 14h32m40s & +36d18m08s & 0.013098 & Sa \\
            UGC\,09711 & 2 & 15h06m37s & +09d26m19s & 0.027915 & Sab \\
            NGC\,6394 & 2 & 17h30m21s & +59d38m24s & 0.028471 & Sbc \\
            NGC\,6762 & 2 & 19h05m37s & +63d56m03s & 0.009729 & Sab \\
            UGC\,12348 & 2 & 23h05m19s & +00d11m22s & 0.025097 & Sb \\
            \hline
        \end{tabular}
        \begin{tablenotes}
            \item[1]The complete data is published in an online catalog at \url{http://132.248.1.15:8001/CALIFA_AGN_hosts.csv}.
        \end{tablenotes}
    \end{threeparttable}
    \label{tab:final_sample}
\end{table*}

We base our selection of Type-I AGN hosts on the presence of a broad component of \Ha in the central spectra. With {\sc Fit3D} \citep{SFS.etal.2016} we perform a fit of the spectral region around the \nii doublet and \Ha with a model that comprises four gaussians: (i) a broad component of \Ha (FWHM $>$ 1000 km/s), and (ii) three narrow components, the \nii doublet and \Ha itself (FWHM $<$ 300 km/s). Following this process, we select those galaxies with a S/N above 8 for the peak intensity of the measured broad component of \Ha. We adopt the peak intensity and not the integrated flux since when no broad component is present the procedure tends to artificially introduce a nonphysical one with a very low intensity peak. Based on this criterion we selected 8 Type-I AGN candidates. Two more were added after a visual inspection of all the spectra with a broad component intensity with a S/N$>$5, selecting those that clearly show a broad component. Similar visual inspection were performed in previous studies for the same reason (e.g., S18).

For Type-I AGNs we adopted the flux intensities derived from the narrow emission lines modeling performed using the multi-Gaussian fitting for the ratios included in \cref{fig:BPTVO}. For doing so we repeated the same fitting procedure described before for the \Hb+\oiii wavelength range, including a narrow and broad component to model \Hb, just for the AGN candidates. All but 2 of the Type-I candidates fulfill the criteria adopted to select Type-II candidates. The other two may be affected by a inaccurate or imprecise determinations of the narrow component of \Hb, since in both cases it is clearly below the flux intensity of the corresponding broad component. It should be noticed that even in the case where we have a good fit for the broad component of \Ha, the fitting of \Hb is more difficult due to its intrinsically lower flux. Indeed the two Type-I candidates that do not fulfill all the criteria indicated before are above the K01 demarcation curves in at least two of the considered diagnostic diagrams. Besides that, for both AGN Types, we have to take into account that, at least in the considered aperture ($\sim$1\,kpc$^2$), the central ionization may be contaminated by ionizing sources other than AGN, that could affect the observed line ratios. One of these galaxies is NGC\,7469 (lower-right triad of panels at \cref{fig:examples}), a face-on Sc galaxy host of a Type-I AGN with the FWHM of the rms spectrum reaching thousands of km/s \citep[e.g][]{Seyfert.1943, Peterson.and.Wandel.2000} and with \WHacen$\approx$ 100\AA. \citet{Rembold.etal.2017} have shown that the hosts of the strongest AGNs (those with higher \oiii luminosity) present an increasing contribution of young stellar populations at their centre, which could account for the loci of this central spectrum in all BPTVO diagnostic diagrams and the high value of \WHacen at the current spatial resolution of our data.

\begin{figure*}
    \includegraphics[width=\columnwidth]{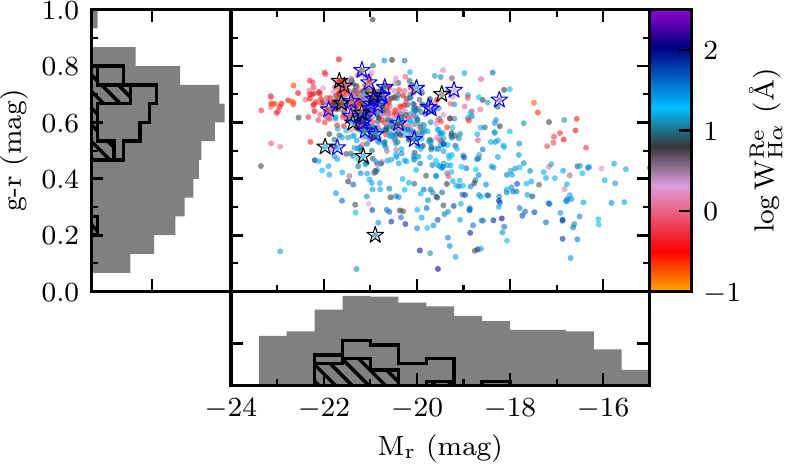}
    \includegraphics[width=\columnwidth]{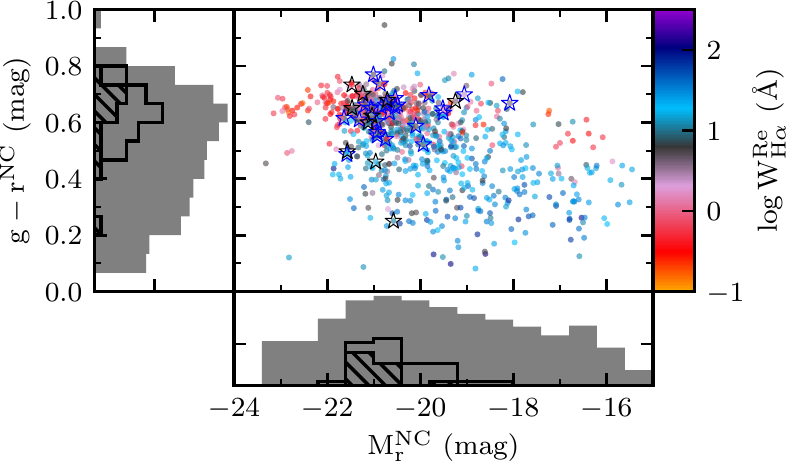}
    \caption{The CMD diagram using the g-r color versus r-band absolute magnitude for the full sample used in this study. The symbols are the same as in \cref{fig:BPTVO} but for the colormap now we choose the value of the EW of \Ha at the effective radius, \WHaRe, following our adopted galaxy classification. The \rpan\ shows the same CMD extracted from the full sample (\lpan) but analysed with the central $3'' \times 3''$ arcsec region subtracted to mitigate the AGN contamination. In both plots we also included the $\log {\rm N}$ histograms of each parameter with colors representing: all sample (grey), All AGN hosts (open black) and Type-I AGN hosts (hatched black). The color figure can be viewed online.}
    \label{fig:CMD}
\end{figure*}

\subsubsection{Final Sample}
\label{sec:analysis:candidates:sample}

In \cref{tab:final_sample} we present the final sample of candidates to host an AGN, that comprises 34 galaxies (24 Type-I and 10 Type-II). They represent $\sim$4 per cent of the total sample of galaxies. This number is very close to the one reported by S18, using a similar selection criteria for the MaNGA IFS dataset \citep{Bundy.etal.2015}. As pointed out by S18, this kind of selection is biased through galaxies with enough gas to present a clear AGN signature in the optical range of the spectrum. That means that we do not include radio-galaxies (without emission lines), Type-0 objects \citep[e.g.][]{Urry.etal.1995} and very obscured AGNs \citep[e.g.][]{Benn.etal.1998, Hickox.and.Alexander.2018}, in our selection. These are not very usual objects (apart from radio-galaxies) and we believe that they do not impose any strong bias in our sample considering the final goals of our study. In respect to radio galaxies, the involved time-scales are different, thus they could represent distinct evolutionary stages between the radio emission and other AGN activity \citep[e.g.][]{Buttiglione.etal.2010,Tadhunter.etal.2012}. Finally, by construction, we exclude very weak AGNs, as we impose an \WHacen > 3\,\AA\ cut for the AGN host selection. However, it is expected that the effect of this group in the overall evolution of the host galaxy is less important.

When selecting AGN hosts using only the central region ionization, our goal is to classify the bona fide AGN hosts and not all the possible candidates. In this way we do not account for possible displaced active nuclei observed in some mergers \citep[e.g.][]{Menezes.etal.2014} and also recently turned off AGNs (e.g. \citealt{Wylezalek.etal.2018}, using the MPL-5 sample). These authors consider the ionization of all spaxels of a galaxy in their classification of AGN hosts. In this way, as we will discuss ahead in this article, they add galaxies that suffer from ionization by outflows in a great number of spaxels, reaching a high fraction of AGN hosts (303 AGN hosts from 2727 galaxies, $\sim$11 per cent). It is worth noticing that no inclination cut was performed during the selection process. This may introduce a possible issue, since in edge-on galaxies shock ionization due to galactic outflows may mimic the properties of that of AGNs, based on our selection criteria. For example, IC\,2247, assigned here as a Type-II AGN host candidate, is also a candidate to be hosting an outflow in the study of 203 highly inclined ($i > 70^\circ$) galaxies from CALIFA made by \citet{LopezCoba.etal.2019}. In the same line, these authors detected extraplanar gas in IC\,1481, a galaxy classified as Type-I by our analysis. However, it is also worth noticing that they report that both objects present an ionization compatible with AGNs in between a 30 per cent and 60 per cent of their covered FoV. Thus, they have already indicated the possible presence of an AGN. It is important to notice that approximately 30 per cent of our Type-II AGN hosts have $i > 70^{\circ}$ and none of our Type-I AGN hosts are. Thus, inclination does not seem to impose a severe bias in the detection of Type-II AGNs in our sample, since the observed fraction is similar to the one found for highly inclined galaxies.

\section{Results}
\label{sec:results}

\subsection{The properties of the AGN hosts}
\label{sec:results:prop}

Knowing which galaxies host an AGN, we now explore their properties and compare them with those of their non-active counterparts.

First we classify the full sample of galaxies with respect to their stage of SF, separating them in three categories, based on the \WHa at the effective radius, \WHaRe:
\begin{itemize}
  \item Star-forming Galaxies (SFG): \WHaRe > 10\,\AA;
  \item Green-valley Galaxies (GVG): 3\,\AA\ > \WHaRe $\geq$ 10\,\AA;
  \item Retired Galaxies (RG): \WHaRe $\leq$ 3\,\AA.
\end{itemize}
We adopt \WHa as a proxy of the SF stage based on the tight correlation between this parameter and the specific star-formation rate \citep[sSFR; e.g.][]{SFS.etal.2013, Belfiore.etal.2018}. In essence, this classification is similar to the one proposed by \citet{Lacerda.etal.2018} but loosening up the floor limit of SF from 14\,\AA\ to 10\,\AA, since here we are dealing with integrated/characteristic properties of entire galaxies. Additionally, we recall that the \WHaRe is a better proxy of the average \WHa of a galaxy than the actual value derived using a limited aperture spectroscopic exploration. Dividing the sample of 867 galaxies in these three classes we obtain the following distribution: 382 SFGs, where SF process is prominent; 185 GVGs, where a range of different processes could are in place (shocks, jets, AGN, etc), or it does not exist a prevalent process (mixed processes like AGN+SF, HOLMES+SF, AGN+HOLMES, etc); and 300 RGs, where the scant gas is mainly ionized by HOLMES. The strong connection between the mean \WHa, the dominant ionization process and the SF-stage was uncovered and discussed in detail recently by 
\citet{CanoDiaz.etal.2019}.

AGN hosts are clearly more frequent within the GVG class, comprising 18 galaxies (4 Type-I and 14 Type-II). Both SFG and RG classes have 8 AGN hosts each one, with 4 Type-I and 4 Type-II at SFG class and 2 Type-I and 6 Type-II between the retired ones. We will see that this behavior is reflected in most of the properties analysed hereafter. Indeed, this was already noticed, in the exploration of the CMD by \citet{Kauffmann.etal.2003c} and \citet{SFS.etal.2004}, for both Type-II and Type-I AGNs respetively. More recently, \citet{Schawinski.etal.2014} replicate both results in more detail. Similar results were reported at a wide range of redshifts by different authors \citep[e.g.][]{Silverman.etal.2008, Wang.etal.2017}.

\begin{figure*}
    \includegraphics[width=\columnwidth]{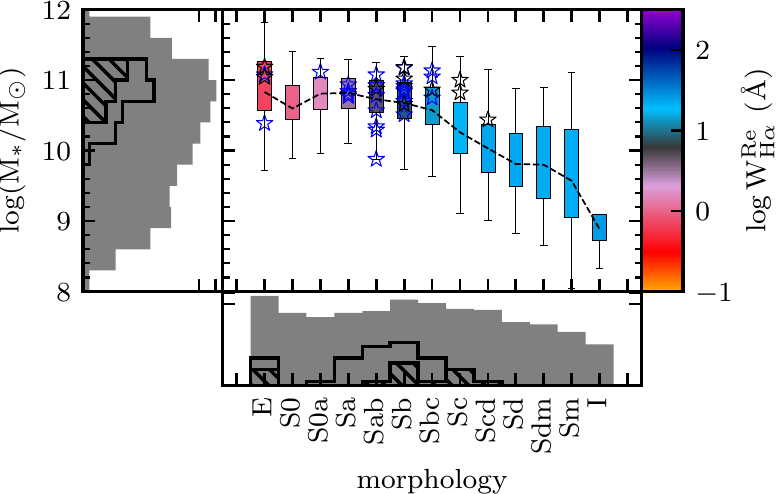}
    \includegraphics[width=\columnwidth]{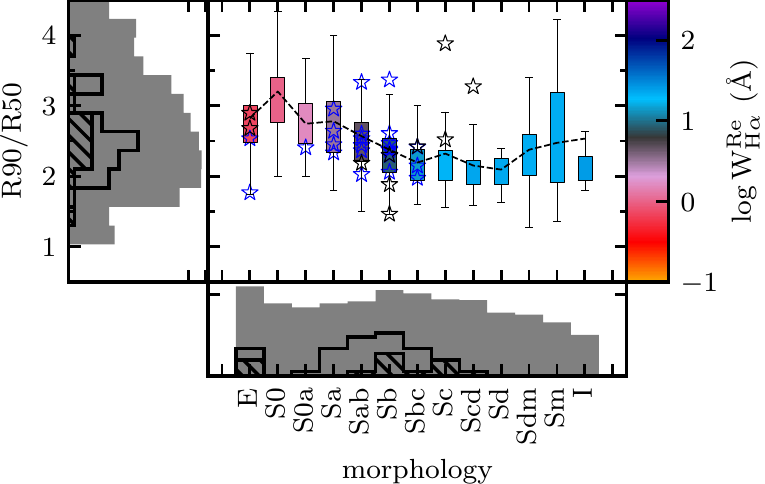}
    \includegraphics[width=\columnwidth]{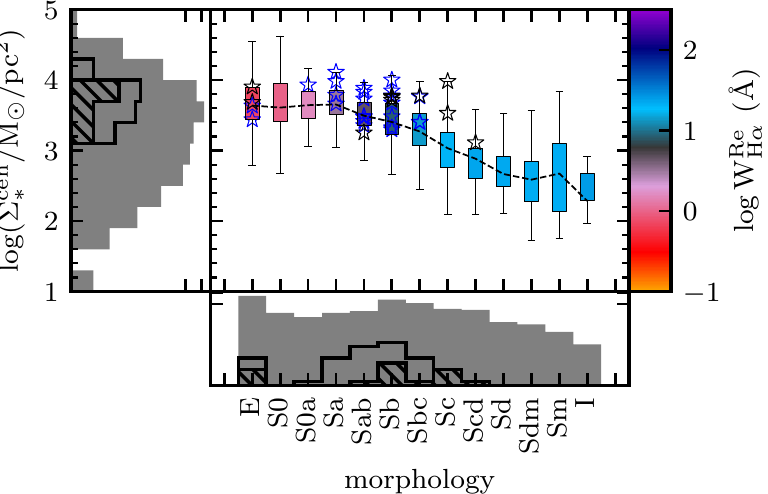}
    \includegraphics[width=\columnwidth]{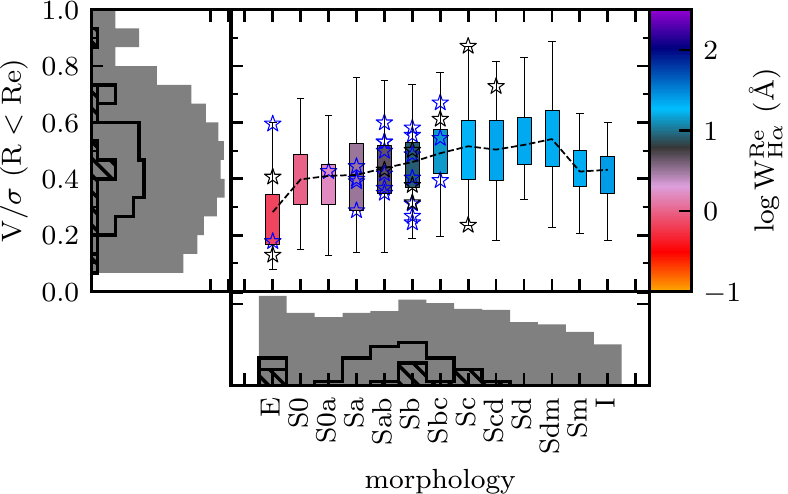}
    \caption{Distribution of box plots of different stellar properties along the morphological type of the galaxies. The boxes are colored by the mean \WHaRe, for each morphological class.
    \ULpan: integrated stellar mass, ${\rm M}_\star$.
    \URpan: concentration index, R90/R50.
    \LLpan: central stellar mass surface density, $\Sigma_\star^{\rm cen}$.
    \LRpan: the ratio between velocity and velocity dispersion, V/$\sigma$, within one effective radius.
    In each panel, the dashed line represents the mean of the distribution on each morphological bin. Symbols and histograms are the same as in \cref{fig:BPTVO} and \cref{fig:CMD} respectively. The color figure can be viewed online.}
    \label{fig:prop_morph}
\end{figure*}

\subsubsection{Color-Magnitude diagram}
\label{sec:results:prop:CMD}

Galaxies present a clear correlation between their colors and their morphology, as already highlighted in the seminal article from \citet{deVaucouleurs.1961}. In a more recent study, \citet{Strateva.etal.2001} verified this bimodality in the $u-r$ color for $\sim$148,000 galaxies from the SDSS survey. The bimodality is more evident in the CMD, showing a density peak for what we know as the red sequence population, dominated by early-type galaxies with dominant old-stellar populations and a blue cloud populated by late-type star-forming galaxies, with younger stellar populations \citep{Bell.etal.2004}. The luminosity of a source correlates very well with its mass \citep[via the M/L ratio that correlates with the color, e.g.,][]{Bell.and.deJong.2001}. Thus, red and old population galaxies are in general more massive and the blue and young ones are the less massive.

The leftmost diagram of \cref{fig:CMD} show the CMD for our sample. We use the $g-r$ color vs. the $r$-band absolute magnitude, M${}_{\rm r}$, in order to verify the location occupied by the AGN hosts in this plane. For the total sample, the bimodality is evident, with the two denser regions indicated before clearly visible: the red sequence and the blue cloud. The region in between these two more populated places is the so-called {\it green valley}. Those galaxies corresponds exactly with those selected using the \WHaRe criterion indicated before. The lack of galaxies in the GV is interpreted as a fast evolution between SFGs and RGs, i.e., a quick halting of the SF (accompanied by a dramatic change in the morphology, structural properties and dynamics of galaxies) compared to the Hubble time. The fraction of GVG in this work is $\sim$20 per cent of the total, the lowest population with respect to the other two groups.

The mean values of the $g-r$ color and M${}_{\rm r}$ are 0.47 mag and -19.46 mag for the SFG, 0.60 mag and -20.46 mag for the GVG and 0.65 mag and -20.84 mag for the RG, respectively. The $g-r$ color seem to be less efficient to segregate between GVGs and RGs \citep[e.g.][]{Kauffmann.etal.2007} that other colors involving UV-bands or other parameters, like the sSFR or the \WHa. The \WHaRe (color-code used in \cref{fig:CMD}) indicates us that the red sequence is populated by galaxies which main ionization is compatible with post-AGBs. On the other hand, galaxies in the blue-cloud present higher \WHaRe values, therefore they are compatible with ionization by SF. As indicated before, several works found AGN hosts populating the GV in the UV-optical CMD too \citep[e.g.][and references therein]{Martin.etal.2007, Salim.etal.2007, Salim.etal.2014}. The AGN hosts of our sample are located in between the two most populated regions of the CMD plane, found at the bluer part of the red sequence, in agreement with former works (e.g. \citealt{Kauffmann.etal.2003a, SFS.etal.2004}; S18). However, regarding their colors, AGN hosts are difficult to differentiate from galaxies mostly ionized by post-AGBs. The $g-r$ colors of Type-I and Type-II AGN hosts are $\sim$0.60 and $\sim$0.65 mag, respectively, and the mean value for all AGN hosts is $\sim$0.63 mag. Thus, a value in between the GVG and RG values. Regarding the M$_{\rm r}$, the mean value for AGN hosts is approximately the same than that of RGs, with -21.18 mag for Type-I and -20.68 mag for the Type-II hosts.

AGNs could present a high luminosity compared with their host galaxies, over-shining them, and dominating the overall colors. Thus, they may change the galaxy position at the CMD. For Type-I AGN hosts this effect could be even stronger \citep[e.g.][]{SFS.etal.2003, SFS.etal.2004, Jahnke.etal.2004a, Jahnke.etal.2004b, Zhang.etal.2016}. With the current dataset we repeat the experiment already done by S18 masking the central $3'' \times 3''$ region from the datacubes and recreating the CMD in order to evaluate the possible AGN contamination upon the colors. We see in the \rpan\ of \cref{fig:CMD} this experiment. In agreement with S18, we do not find any significant difference between both diagrams. In summary, Type-I hosts shift, in average (standard deviation), -0.01 mag (0.02 mag) in the $g-r$ axis and 0.25 mag (0.08 mag) in the M${}_{\rm r}$ one, while Type-II hosts shift by -0.02 mag (0.01 mag) in the $g-r$ axis and 0.18 mag (0.07 mag). In both cases the change is negligible. Therefore, the position of AGNs in the transition regime between the red sequence and the blue cloud is not produced by the nuclear source itself, as excluding the central regions neither the positions of Type-I AGN hosts nor those of the Type-II AGN hosts change significantly.

\subsubsection{The morphology}
\label{sec:results:prop:morph}

\begin{table*}
    \centering
    \caption{Distribution of galaxies in morphological classes, highlighting the AGN hosts. We include the average values of properties in order to facilitate evaluation and comparison. We do not find optical AGNs in the center of late-type disk galaxies (> Scd).}
    \begin{threeparttable}
        \begin{tabular}{c|c|c|c|c|c|c|c|c}
            \hline
            M. type & N (per cent) & all AGN\tnote{1} & Type-I\tnote{2} & Type-II\tnote{2} & $\log({\rm M}_\star$/M$_\odot)$ & R90/R50 & $\log(\Sigma_\star^{\rm cen}/{\rm M}_\odot/{\rm pc}^2)$ & V/$\sigma$ (R < Re) \\
            \hline
            \hline
            E & 163 (19) & 4 (12;2) & 2 (6;20;1) & 2 (6;8;1) & 10.83 & 2.82 & 3.64 & 0.28 \\
            S0 & 59 (7) & -- & -- & -- & 10.60 & 3.20 & 3.61 & 0.40 \\
            S0a & 46 (5) & 1 (3;2) & -- & 1 (3;4;2) & 10.81 & 2.73 & 3.64 & 0.41 \\
            Sa & 59 (7) & 4 (12;7) & -- & 4 (12;17;7) & 10.82 & 2.78 & 3.65 & 0.41 \\
            Sab & 66 (8) & 8 (24;12) & 1 (3;10;2) & 7 (21;29;11) & 10.73 & 2.57 & 3.50 & 0.44 \\
            Sb & 130 (15) & 10 (29;8) & 3 (9;30;2) & 7 (21;29;5) & 10.68 & 2.38 & 3.41 & 0.46 \\
            Sbc & 106 (12) & 4 (12;4) & 1 (3;10;1) & 3 (9;12;3) & 10.57 & 2.19 & 3.27 & 0.49 \\
            Sc & 75 (9) & 2 (6;3) & 2 (6;20;3) & -- & 10.26 & 2.32 & 3.03 & 0.52 \\
            Scd & 71 (8) & 1 (3;1) & 1 (3;10;1) & -- & 10.03 & 2.15 & 2.88 & 0.50 \\
            \hline
            Sd & 34 (4) & -- & -- & -- & 9.81 & 2.09 & 2.67 & 0.52 \\
            Sdm & 30 (3) & -- & -- & -- & 9.80 & 2.38 & 2.59 & 0.54 \\
            Sm & 19 (2) & -- & -- & -- & 9.57 & 2.47 & 2.67 & 0.43 \\
            I & 9 (1) & -- & -- & -- & 8.89 & 2.53 & 2.29 & 0.43 \\
            \hline
        \end{tabular}
        \begin{tablenotes}
            \item[1] We show the number of galaxies, the percentage over the total of objects at the considered column (all AGN hosts, only Type-I hosts and only Type-II hosts) and over the total present of each morphological class. E.g. We have a total of 9 AGNs among Sab galaxies. They represent 24 per cent of all AGNs and 12 per cent of all Sab.
            \item[2] We present the same statistics of all AGN with the inclusion of the percentage over the total number of each AGN Type. E.g., we have a total of 34 AGNs and 10 of them are among Sb galaxies. Three of them are classified as Type-I (9 percent of all AGNs, 30 per cent of all Type-I hosts and 2 per cent of Sb galaxies); and 7 are Type-II AGN hosts (21 percent of all AGNs, 29 per cent of Type-II, and 5 per cent of Sab galaxies).
        \end{tablenotes}
    \end{threeparttable}
    \label{tab:morph}
\end{table*}

We continue our study comparing representative physical properties that can elucidate which type of galaxies are hosting an AGN and which are the differences with respect to those that are not (if any). \cref{fig:prop_morph} shows the distribution of the integrated stellar mass, R90/R50 concentration ratio, central stellar mass surface density and the velocity-to-velocity dispersion ratio, V$/\sigma$, along morphological classes. In general, the average distribution
of such properties follows the expected values for the majority of galaxies at each class. Later-types show smaller total stellar-mass, are more concentrated in light and in stellar-mass surface density with its stars following less disturbed orbits, in comparison to early-type ones (more massive, concentrated and usually supported by random orbits).

\begin{figure*}
    \includegraphics[width=\columnwidth]{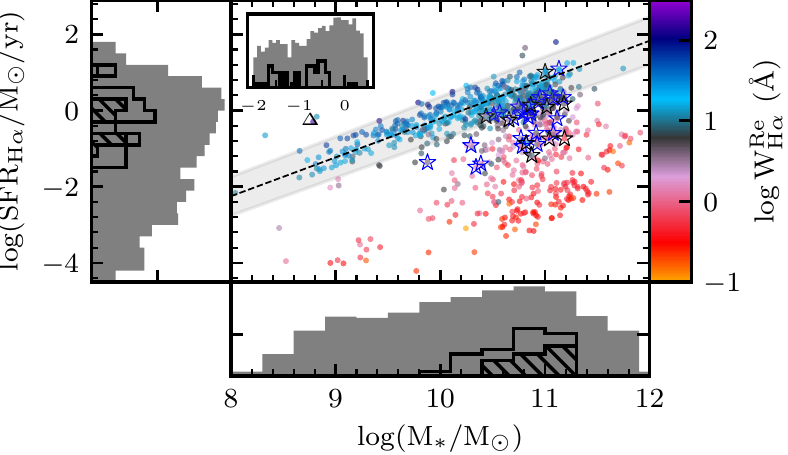}
    \includegraphics[width=\columnwidth]{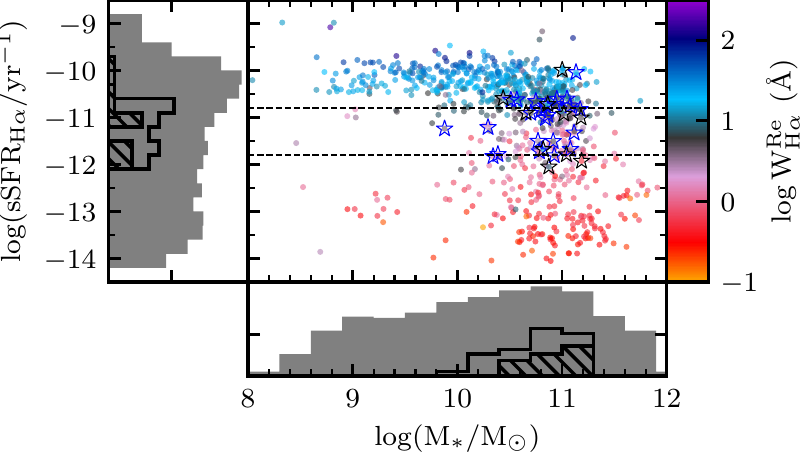}
    \caption{Integrated star-formation rate (SFR) and integrated specific star-formation rate (sSFR) versus stellar mass for all sample used in this study. \Lpan: The SFR from a linear transformation from the integrated \Ha luminosity decontaminated for diffuse emission, as the contribution of other ionization sources plays an important role for the integrated \Ha values (cf. \cref{sec:analysis:syntheml}). We should take in account that for the RGs these values have to be considered as an upper-limit to the real SFR. The black dashed line represents the orthogonal distance regression (ODR), averaged by the error in both properties, for the so called star-formation main sequence (i.e., the loci of SFGs; c.f. \cref{eq:SFMS}) with the filled area representing $\pm 1\sigma$ interval. In the inset panel we add the histogram of the $\Delta$($\log$SFR), the distance of each galaxy to the SFMS defined by \cref{eq:SFMS}. \Rpan: The integrated sSFR (SFR/${\rm M}_\star$) versus the stellar mass. The two dashlines ($-11.8 > \log({\rm sSFR}) > -10.8$) delimits the interval for transitional galaxies in local universe by \citet{Salim.etal.2007}. The symbols, colors and histograms are the same as those presented in \cref{fig:BPTVO}, \ref{fig:CMD} and \ref{fig:prop_morph}. The color figure can be viewed online.}
    \label{fig:SFMS}
\end{figure*}

If morphology is somehow bound to the evolution of galaxies, and AGNs play a role in the quenching process, they should be located as transition objects between spirals (SFGs) and ellipticals (RGs). It is clearly appreciated in all panels in \cref{fig:prop_morph} that the AGN hosts are indeed in the spiral-to-elliptical transitory space. Table \ref{tab:morph} summarize the distribution of galaxies in morphological classes and how the AGN hosts spread among them, including the average values of every property at each morphological class. We see a predilection to Sab-Sb galaxies harbouring an AGN while none are found in later-types (> Scd). Our sample of AGN hosts contains 4 early-type elliptical galaxies (E), 29 from early to late spirals (>= Sa) but only one among lenticular ones (S0+S0a). Statistically, we found $\sim 12$ per cent of all Sab galaxies (8 out of 66) hosting an AGN. Finally, we have 10 hosts among the Sb galaxies, but those represent only $\sim 8$ per cent of this class galaxies (10 in 130).

In the \ulpan\ of \cref{fig:prop_morph} we see that the AGN hosts are strongly biased toward massive galaxies when comparing with the non-active ones, with their stellar masses peaking at ${\rm M}_\star = 10^{10.8}\,{\rm M}_\odot$. Indeed, they are distributed mostly at the massive end at each morphology bin, except among E and Sab galaxies, where they spread along the full range of masses for this class. In total, $\sim$74 per cent of the AGN hosts are above the mean stellar mass distribution. For Type I the effect is stronger, with 90 per cent above the mean value, although 67 per cent of Type-II are also above the mean. This behavior is similar regarding the central stellar mass surface density ($\Sigma^\star_{\rm cen}$), with $\sim$80 per cent of the AGN hosts above the mean value, hence more centrally peaked, with a mean value of $10^{3.7}\,{\rm M}_{\odot}/{\rm pc}^2$.

The main trend of the concentration index (R90/R50) follows an anti-correlation from late to early-to-late type galaxies (with higher values for the later ones). The mean value for the AGN hosts is 2.6, identical the value found to separate early from late-types \citep[e.g.][]{Shimasaku.etal.2001, Strateva.etal.2001}. This means that the AGN hosts are also in between classes regarding the concentration index too.

Regarding the $V/\sigma$ ratio within the effective radius (\lrpan\ in \cref{fig:prop_morph}), galaxies in general follows the expected trend. Early-type galaxies, supported in average by disordered motions (pressure), present smaller values of this ratio. Conversely, later-type ones (up to Sd/Sdm) present larger and large values \citep[e.g.][]{Cappellari.2016,Blanton.and.Moustakas.2019}. AGN hosts follow the main distribution too. Conversely, they are confined to morphological types with a substantial fraction of the hot/warm orbits \citep{Zhu.etal.2018a, Zhu.etal.2018b}.

\subsubsection{The integrated SFR-M${}_\star$ diagram}
\label{sec:results:prop:SFMS}

\cref{fig:SFMS} shows the distribution of the SFR along the M$_\star$ for the galaxies in our sample. The distribution shows the bimodal behavior already discussed in previous articles \citep[e.g.][]{Brinchmann.etal.2004, Daddi.etal.2007, Salim.etal.2007, Renzini.and.Peng.2015}, with two sequences, one for SFGs (the so-called star-formation main sequence, SFMS), and another one for RGs. The number of galaxies in between these two sequences is very scarce, being larger for more massive galaxies than for less massive ones. Spectral energy distribution (SED) fit including UV-to-IR theoretical models unveil that our current derivation of the SFR for RGs (i.e., at very low SFRs) is overestimated \citep{Bitsakis.etal.2019}. \citet{LopezFernandez.etal.2016a} achieved the same conclusion using the stellar synthesis code \starlight\ \citep{CF.etal.2005} updated for UV photometric data. Despite all uncertainties in the process, the \Ha hDIG decontamination implemented here should correct somehow this overestimation. In any case, its effect would be to decrease the bimodality, pushing up RGs in this diagram towards the loci of SFGs.

Moreover, the \lpan\ of \cref{fig:SFMS} presents an orthogonal distance regression (ODR) for the SFGs with the filled area representing $\pm 1\sigma$ interval. This regression minimizes the orthogonal distance from each point to the average distribution fitted by a model, in this case, a linear one. The same process is performed for all regressions along this article. The process takes into account the errors in both parameters. As indicated in \cref{sec:results:prop}, apart of the differences in the selection of SFGs (since we define a GVG class in between the RG and SFG classes), a bigger sample with a different regression method leads to distinct derived slope and zero-point comparing with recent results \citep[e.g.][]{CanoDiaz.etal.2016}. Our best-fit values for the SFMS is:
\begin{equation}
  \log({\rm SFR}_{\Ha}) [{\rm M}_{\odot}/{\rm yr}] = 1.02_{\pm 0.03} \log {\rm M}_\star [{\rm M}_{\odot}] - 10.41_{\pm 0.27}.
  \label{eq:SFMS}
\end{equation}
\noindent The dispersion in the fit is $\sigma_{\rm ODR}=0.15$ dex and the one projected to the SFR-axis is 0.22 dex and the Spearman's Rank correlation coefficient for the distribution is $r=0.87$.

\begin{figure*}
    \includegraphics[width=\columnwidth]{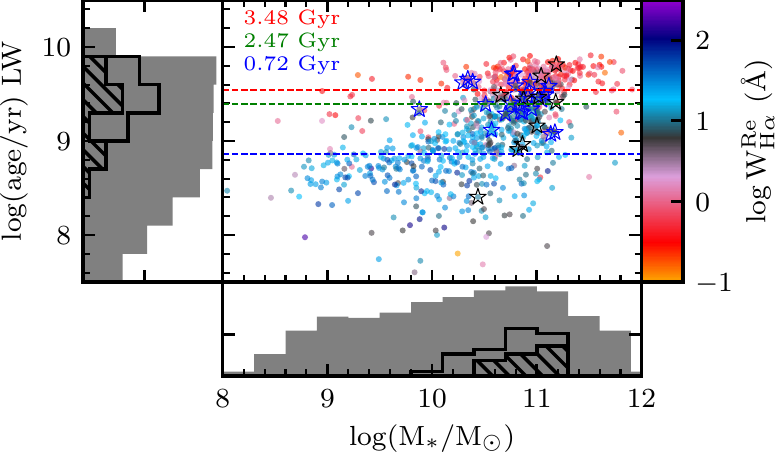}
    \includegraphics[width=\columnwidth]{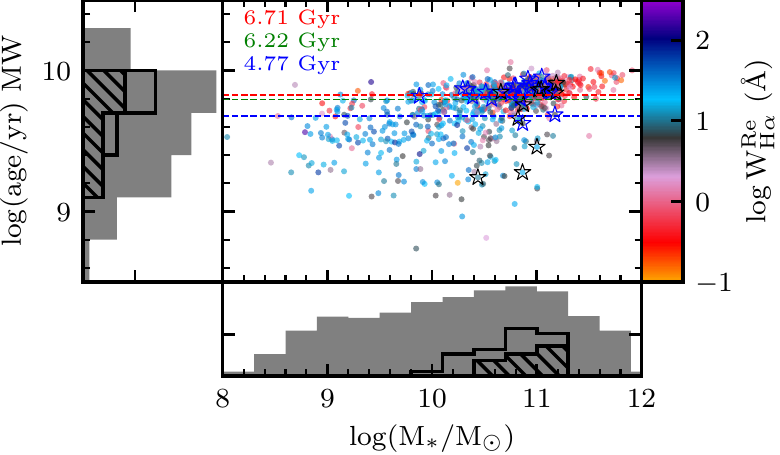}
    \caption{Distribution of stellar age at the effective radius versus the stellar mass for our sample of galaxies. \Lpan: The light-weighted mean stellar age. \Rpan: The mass-weighted mean stellar age. The three dashed lines represent the average values (shown at the top-left position of each panel) for SFGs (blue), AGN hosts (green) and RGs (red). The symbols, colors and histograms are the same as those presented in \cref{fig:BPTVO}, \ref{fig:CMD} and \ref{fig:prop_morph}. The color figure can be viewed online.}
    \label{fig:M_t}
\end{figure*}

Concerning present-day optical AGN hosts, they occupy mostly the right side of the plane, in the region between both sequences. This suggests that the AGN have an important role in the SF quench. The position of the AGN hosts in the SFR-${\rm M}_\star$ diagram was reported before by \citet{Shimizu.etal.2015}, \citet{CanoDiaz.etal.2016} and confirmed by \citet{Smith.etal.2016}, \citet{CatalanTorrecilla.etal.2017} and S18. We find both AGN types located most frequently under the SFMS, at the high-mass end of the this relation. Type-I hosts are mostly concentrated at the high-mass end of the this relation (as we see also in the \cref{fig:prop_morph}), however they are distributed along a interval in SFR similar to that for Type-II hosts. This result appears to contradict somehow the one reported by S18 in their MaNGA study. They found that the Type-I are more concentrated in the low-end of the SFMS, and the Type-II are more spread between the low-end of the SFMS and the high-end of the RGs region of the plot. We compute the distance (in log) of each galaxy in the plot to the SFMS, 
i.e. the SFR that they should have if they were average SFGs, as seen in the inset plot at \lpan\ of \cref{fig:SFMS}. They appear to populate the same range of values with the same scatter, with Type-II going a little bit down towards the RGs region. Moreover, only two AGN hosts are above the SFMS, NGC 1667 (Type-II AGN host) and NGC 7469 (Type-I AGN host). The reason could be that both of them are face-on galaxies with a very blue disk. Furthermore, they are considered strong AGNs in our sample (\WHacen > 10\,\AA) and also both exhibit high \WHaRe (above 20\,\AA).

We repeat the same procedure adopted for the analysis of the CMD and remove the central regions of the datacubes to minimize any possible bias in the position of AGN hosts due to a contamination by the AGN itselft. As previously reported by \citet{CatalanTorrecilla.etal.2015}, for Type-II AGN hosts this contamination can be neglected when comparing with the integrated \Ha luminosity. S18 also reported that this conclusion can be extended for the Type-I hosts, indicating that the AGN hosts are indeed in this intermediate region of the SFR-${\rm M}_\star$ plane, between the SFG and the RG, irrespective of the possible contamination of the AGN. We can indeed quantify the offset introduced by this contamination, showing that the mean (stardard deviation) of the SFR changes by 0.03 (0.03) dex for the full sample, 0.06 (0.04) dex for the Type-II AGN hosts and 0.07 (0.03) dex for the Type-I hosts. Indeed, this is well below the typical expected error for these values.

The \Rpan\ of \cref{fig:SFMS} shows the distribution of sSFR along M$_\star$, highlighting exactly this behavior. At high-mass, the sSFR can differentiate better galaxies that have stopped its SF from those that are still actively forming stars than the diagrams shown in previous sections. One more time, AGN hosts are in the region in between (high-mass with lower sSFR) following the colors in the plot, suggesting that the nuclear activity should play a role in halting the SF from the host galaxy. Moreover, high values of sSFR are populated by low (B/T)$_\star$ galaxies and the sequence of dominance of bulge in stellar mass grows as the sSFR decreases, with AGNs present in between them.

\subsubsection{Stellar Mass-Age relation}
\label{sec:results:prop:age_mass}

\begin{figure*}
    \includegraphics[width=\columnwidth]{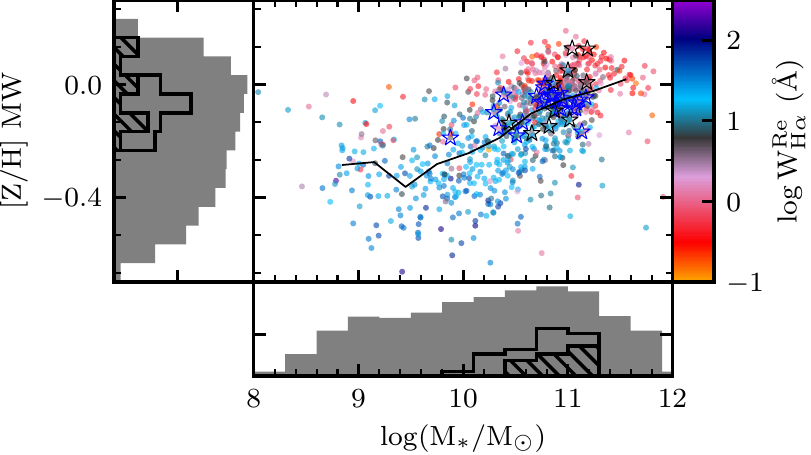}
    \includegraphics[width=\columnwidth]{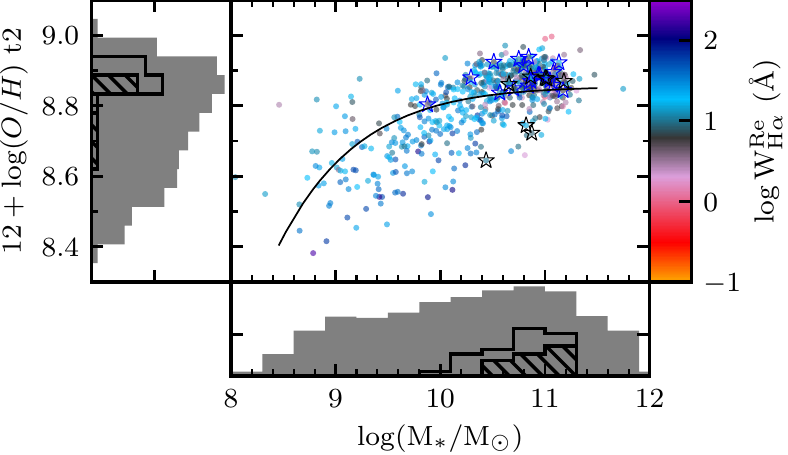}
    \caption{Mass-metallicity relations. \Lpan: Distribution of Mass-weighted stellar metallicity within one effective radius versus the stellar mass for the full sample in this study. The solid line represents the mean value for 0.3 dex bins in stellar mass. \Rpan: Distribution of the oxygen abundance within one effective radius using the $t2$ calibrator, versus the stellar mass for galaxies in the sample that show gas emission lines compatible with being ionized by SF. The oxygen abundance estimation is described at \citet{SFS.etal.2017}. The symbols, colors and histograms are the same as those presented in \cref{fig:BPTVO}, \ref{fig:CMD} and \ref{fig:prop_morph}. The color figure can be viewed online.}
    \label{fig:MZR}
\end{figure*}

In \cref{fig:M_t} we show the distributions along stellar mass of the luminosity weighted (LW, \lpan) and mass-weighted (MW, \rpan) mean stellar age evaluated at the effective radius. Ages are derived based on the stellar population synthesis described in \cref{sec:analysis:syntheml}. Furthermore, we calculate the mean age for SFGs (blue dashed line), RGs (red dashed line) and AGN hosts (green dashed line), respectively. The evolution of the distribution is evident to the eye. In both panels, the AGN host population has mean stellar ages in between SFGs and RGs. The LW mean stellar age highlights better the young population, which are dominant in light. But the old population is predominant in number (and mass), so in the MW version, the old stars are dominating the average. With this analysis we can conclude that the predominant stellar population in AGN hosts galaxies is old, since the MW mean stellar age is close to that of the RGs and with an intermediate young stellar population. We see a very good agreement with hyphothesis that the AGN phase is part of an evolutive sequence in between both dominant behaviors (star-forming and retired classes). This result totally agrees with the one derived from the exploration of the SFR-M$_\star$ diagram, being based on an independent analysis.

In this regards it is important to compare the time required for a galaxy to evolve from SF to retired (using the age difference of the stellar populations as a proxy) with the known life-time of the nuclear activity. Studies on AGN life-time agree that it lasts between $10^5$ to $10^9$ years, depending what is consider to constrain the AGN activity (ratio between the number of RG/GV/SFG populations, period of X-ray being emitted by a central event, length of radio jets, statistics in function of redshift, etc). \citet{Tadhunter.etal.2012} studied a dual radio-loud/radio-quiet AGN system. Through the assumption that both AGN were triggered at same time by a major galaxy merge, they observe radio emission from hot spots away from the nucleus even not observing emission from the narrow-line region of one of the AGNs. Deriving the life-time of the radio emission episode, they conclude that the AGN may be switched in the last $\sim10^6$\,yr. More recent, \citet{Stasinska.etal.2015b} using statistical arguments regarding the local (z < 0.4) population of galaxies with a {\it detectable} AGN and with M$_\star$ > $10^{10}$ M$_\odot$ proposed an upper-limit for the AGN life-time of about 1-5\,Gyr. From a sample of AGN hosts with X-ray observations, \citet{Schawinski.etal.2015} conclude that the AGN duty-cicle is composed by a sequence of various short-lived ($10^5$\,yr) processes of mass being accreated to the galaxy centre (feeding the BH).

Being various rapid processes or a long one, the observed time of an AGN is various orders of magnitude shorter that the Hubble time, so this kind of process is not present during all the galaxy life-time (if nuclear activity is present in all galaxies). On the other hand, we find that the stellar population of AGN hosts are in average 1\,Gyr younger than RG galaxies and about 2\,Gyr older than SFG regarding LW mean stellar age. This interval between RG and AGN hosts decreases when adopting the MW stellar ages, but the outcome is the same. S18 did the same exploration using data of the MaNGA survey. Despite of the quantitative differences in the mean stellar ages for different classes we reach the same qualitative conclusions. We attribute the quantitative differences mainly to different spatial resolution and redshift range covered by the CALIFA and MaNGA samples.

As indicated before the population of AGN hosts are concentrated among the GVGs, but they comprise only $\sim$10 per cent of all these galaxies. So, either the nuclear activity of the other 90 per cent of GVG is too weak for our detection criteria or the differences in the time-scales between the halting of the SF and the AGN activity is of a factor $\sim$10 larger (i.e., only 1/10th of GVGs would be seen in the AGN period).

\subsubsection{Mass-Metallicity Relation}
\label{sec:results:prop:age_met}

The \lpan\, of \cref{fig:MZR} shows the distribution of the MW stellar metallicity at the effective radius versus the stellar mass for all sample. The clear correlation seen in the plot is known as the stellar mass-metallicity relation \citep[MZR;][]{Gallazzi.etal.2005,ValeAsari.etal.2009,GonzalezDelgado.etal.2014b}. Like in the case of the stellar ages, metallicities are derived based on the stellar population analysis described in \cref{sec:analysis:syntheml}. The stellar MZR is steeper than the nebular one, for the same reasons that we obtain flatter stellar metallicity profiles when averaging only over young stars \citep{GonzalezDelgado.etal.2014b}. The AGN hosts are concentrated in the high-mass, high-metallicity corner of the distribution, and mostly ($\sim$60 per cent) above the mean distribution. This result holds for both Type-I and Type-II AGN hosts.

\begin{figure*}
    \includegraphics[width=\columnwidth]{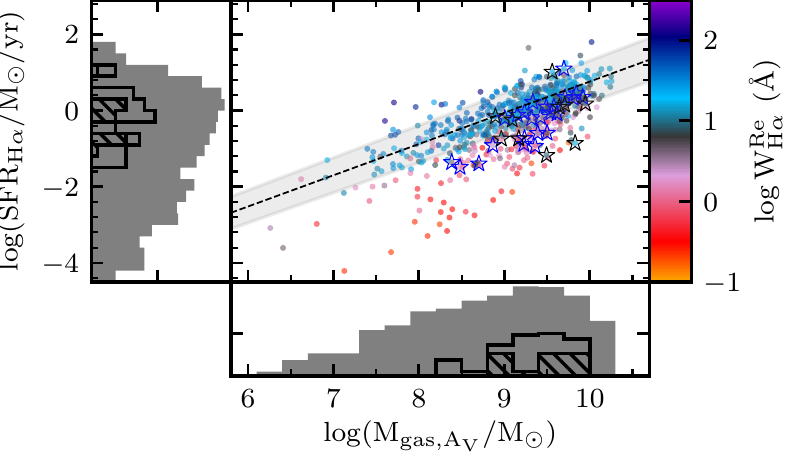}
    \includegraphics[width=\columnwidth]{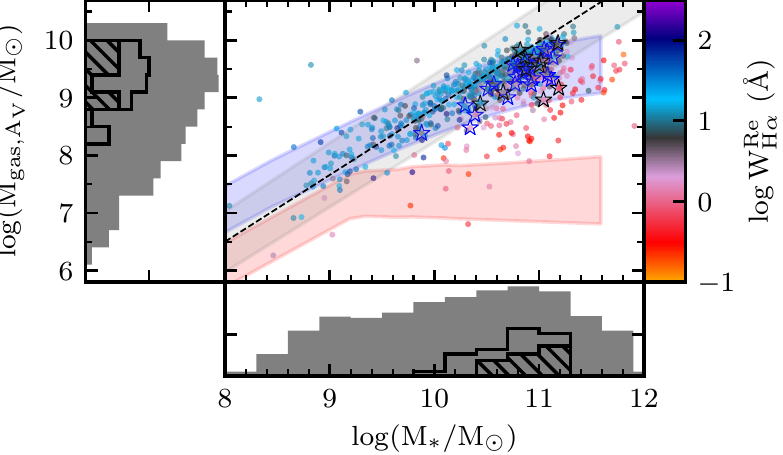}
    \caption{\Lpan: Distribution of the integrated SFR along the estimated integrated molecular gas mass. \Rpan: Integrated gas mass versus the stellar mass. Here, both panels show the ODR for SFGs in the sample, always considering errors in both axis. The filled area represents the $\pm 1\sigma$ interval around the fit. The dashed blue and red filled areas at the \rpan\ are the mean distributions for late- and early-type galaxies respectively, from Rodriguez-Puebla et al (submitted), delimited by the corresponding $\pm 1\sigma$ intrinsic scatter around the distributions. The symbols, colors and histograms are the same as those presented in \cref{fig:BPTVO}, \ref{fig:CMD} and \ref{fig:prop_morph}. The color figure can be viewed online.}
    \label{fig:SK_MMgas}
\end{figure*}

At the \rpan\ of \cref{fig:MZR} we show the gas-phase version of the MZR for the 587 galaxies with emission lines detected fulfilling the criteria to derive the oxygen abundance at the effective radius following \citep{SFS.etal.2014} and successive studies \citep[e.g.][]{SanchezMenguiano.etal.2016, SFS.etal.2017, BB.etal.2017}. The adopted proxy for nebular metallicity is the $t2$-calibrator for the oxygen abundance (O/H), although similar qualitative results are found when adopting any other strong-line calibrator \citep[e.g.][]{SFS.etal.2017}. We derived the best fitted MZR using the formula described in \citet{SFS.etal.2013} (included in the \lpan\ of \cref{fig:MZR}). Of the total number of AGNs, 25 are present in this plot and 21 (84 per cent) of them are above the mean distribution. All but one Type-II are above the mean distribution, and 5 out of 8 Type-I too.

For the AGN hosts, the main distribution along the MZR diagram seems to follow a different trend than the one found by S18 for the MaNGA galaxies. There AGN hosts were found mostly under the mean trend for the entire sample. Assuming that we have the same drawbacks in measuring metallicity of AGN hosts, the differences in mass and redshift range of the samples appear to be the main reason. MaNGA samples a slightly wider range of stellar masses, in particular covering a mass range that goes up to $10^{12}\ {\rm M}_\odot$. Due to the redshift cut of the CALIFA sample, it is difficult to include such massive galaxies (that are very limited in number at low redshift). A detail inspection of the right panel of the Figure 6 in S18 tell us that the AGN fraction tops at very high stellar masses, and all AGN hosts above $10^{11}\ {\rm M}_\odot$ have estimated oxygen abundances below the mean value. Once removed those massive AGN hosts, not observed by CALIFA, both results are pretty consistent.

Moreover, metallicity increases as the Universe evolves. Thus higher-redshift galaxies (e.g., MaNGA) present a slightly lower metallicity (see Figure 6 in \citealt{LaraLopez.etal.2013} and also Figure 1 in \citealt{Zahid.etal.2014a}). Thus, there is an anticorrelation between the mean oxygen abundance (at a fixed stellar mass) and the redshift. Indeed, we found that the median redshift for the AGN hosts is lower than for non-active galaxies at this mass range, what may give a possible explanation for the observed (aparent) discrepancy with previous results (e.g. S18).

Looking from another perspective, \citet{Thomas.etal.2019} found a systematic positive offset in the oxygen abundances for Seyfert galaxies in comparison to SFGs from the SDSS DR7. They inspect many possible methodological causes that could artificially create this offset. However, they did not not reach a denifitive conclusion. Nevertheless, they proposed that top-heavy IMFs in the accretion-disk may produce an increase in abundance. The offset found in our study is much smaller. Selecting galaxies with stellar masses above $10^{10}\ {\rm M}_\odot$, we find the AGN hosts to be just 0.01 dex above the mean value and about 0.03 dex above the corresponding value for SFGs in the same mass range. Those differences are far too small compared with the accepted errors for oxygen abundance calibrators.

In summary, regarding the MZR, and in particular the stellar one, AGN hosts seem to be in the transition between the location of SFGs and RGs in this diagram. Again, this reinforces the results from previous sections suggesting that AGNs are somehow connected with the quenching process.

\subsubsection{The gas content in AGN hosts}
\label{sec:results:prop:gas}

So far all our results have made clear that the AGN hosts are mostly transitional objects (GVG) with integrated properties in between those of SFGs and RGs. However, this does not clarify which is the mechanism required to halt the SF nor if it is an unique process. We already indicated that the main mechanisms proposed for quenching involves either removal or heating of the molecular gas. We explore in this section the gas content on galaxies and its connection with the presence of an AGN.

\cref{fig:SK_MMgas} \lpan\, shows the scaling relation between molecular gas and SFR, an integrated version of the SK-law/relation \citep[][]{Schmidt.1959,Kennicutt.1998}. The SK-law was originally defined only for SFGs, and using intensive quantities (surface densities) of the involved parameters. The $\Sigma_{\rm gas}$ in the original SK-law is referred as the total gas surface density ($\Sigma_{\rm HI}$ plus $\Sigma_{\rm H_2}$), but here we parametrize our relation only in terms of the molecular gas, which traces the SFR better than the total gas \citep[e.g.][]{Wong.and.Blitz.2002, Bigiel.etal.2008}. There is a clear tight relation that tell us that the SFR is directly proportional to the surface density of the molecular gas. This relation holds to kpc-scales regarding the molecular gas as a manifestation of the spatial and time averaging of local physical processes \citep[e.g.][]{Kennicutt.and.Evans.2012}. The  slope of the SK-law ranges from 1 to 3 in the literature (e.g. \citealt{Gao.and.Solomon.2004, Bigiel.etal.2008, Narayanan.etal.2012, Leroy.etal.2013}; S18)   , being 1.4 the nominal one derived by \citet{Kennicutt.1998} for starburst galaxies. Those values are in agreement to the physically motivated range, that goes from 0.7 to 2 depending of the considered densities, timescales and physical details of the SF process \citep{Bigiel.etal.2008}. Quantitative differences in the derived slopes in the literature are attributed to resolution of the data, adopted IMF, considered gas (total or molecular) probed by different tracers and conversion factors (e.g. CO, HCN, dust-to-gas ratio), and differences in the SFR tracers (e.g., \Ha, IR, UV, multi-SSP or multi-wavelenght analysis) (S\'anchez et al. submitted).

Comparing with the values derived by S18, we have differences in the apertures for which we derive the SFR and M$_{\rm gas}$, a higher value of \WHa adopted to select SFGs and a different method to perform the regression. We find a very tight correlation ($r=0.85$) with a steeper slope for the SFGs. The best fit using the ODR method to the data is:
\begin{equation}
  \log({\rm SFR}_{\Ha}) [{\rm M}_{\odot}/{\rm yr}] = 0.82_{\pm 0.03} \log {\rm M}_{\rm gas} [{\rm M}_{\odot}] - 7.44_{\pm 0.25},
  \label{eq:SK}
\end{equation}
\noindent with dispersions $\sigma_{\rm ODR}=0.22$ dex and $\sigma_y=0.28$ dex. For the entire sample, the relation is steeper but less concentrated (the slope is 0.93, the zero point, $-8.65\,{\rm M}_{\odot}/{\rm yr}$, $\sigma_{\rm ODR}=0.33$ dex, $\sigma_{\rm ODR}=0.46$, $r=0.70$) that the one reported by S18. The bimodal behaviour observed in most of the previous analysed diagrams (CMD, SFR-M$_\star$, age-M$_\star$), but not in all of them (MZR), is not so prominent in here neither. There is an offset in the SFR between SFG and RG at a given M$_{\rm gas}$, that increases for M$_{\rm gas}<10^{9.5}$ M$_\odot$, a result that was not found by S18 too. As we explain in \cref{sec:results:prop:SFMS}, our hDIG-corrected SFR$_{\Ha}$ results in lower values of SFR for hDIG dominated galaxies, which are mostly RG with low gas fractions. Thus, adding GVGs and RGs to the linear fit of the relation increases the slope, zero-point (as we are adding high $\tau_{\rm dep}$ galaxies) and the scatter, decreasing the correlation coefficient.

The mean depletion time $\tau_{\rm dep}=\rm M_{\rm gas}/{\rm SFR}$ derived for SFGs is $\approx 1.08$\,Gyr, a value slightly lower than the one found by recent studies \citep[e.g. $\approx$2.2\,Gyr,][]{Leroy.etal.2013, Colombo.etal.2018}. Comparing the SF activity, the SFGs have on average from 0.6 to 1.9 dex higher SFR than RGs at same molecular gas mass, with differences weakely decreasing while increasing M$_{\rm gas}$. Furthermore, to produce the same SFR, RGs needs to have much more molecular gas than SFGs (i.e., the have a lower star-formation efficiency, SFE$=\frac{1}{\tau_{\rm dep}}=\frac{\rm SFR}{\rm M_{\rm gas}}$). Both former results reinforce the idea that SFGs form stars more efficiently. Considering that our estimated SFR for RGs are most probably just upper-limits, the difference in SFE may be even stronger, with RGs have much lower values than what it is reflected in our analysis.

Like in the case of the SFMS, we find here the same AGN hosts (NGC 1667 and NGC 7469)  above the best fit for the SK-law derived for SFGs. On average AGN hosts present lower (higher) SFE than SFG (RG): thus, they are again in between both population of galaxies. They exhibit integrated gas masses over $10^{8.4}$ M$_\odot$ with $10^{9.4}$ M$_\odot$ in average. Therefore they are concentrated in the high-mass end of the plot. Type-I AGN hosts peak at slightly higher masses than Type-II AGN hosts (0.1 dex above in average) with the distribution of M$_{\rm gas}$ having a similar scatter ($\sigma\approx$ 0.4 dex for both). Whether this is a significant result and points towards a real difference between both families of AGNs should be explored using larger samples.

In the \rpan\ of \cref{fig:SK_MMgas} we see the distribution of the estimated molecular gas mass as a function of the integrated stellar mass. The best-fit using the ODR methods for the SFGs is:
\begin{equation}
  \log {\rm M}_{\rm gas} [{\rm M}_{\odot}] = 1.16_{\pm 0.03} \log {\rm M}_\star  [{\rm M}_{\odot}] - 2.78_{\pm 0.29},
  \label{eq:M_Mgas}
\end{equation}
\noindent with $\sigma_{\rm ODR}=0.19$ dex, $\sigma_y=0.29$ dex and a correlation coefficient $r=0.92$.

The RGs present lower values of M$_{\rm gas}$ on average and they cover a wider range of values at same stellar mass than SFGs and GVGs. For SFGs we see a clear tight correlation betwenn M$_{\rm gas}$ and M$_\star$. This points towards the direction that galaxies that have lower gas fractions are those that form stars at lower rates for corresponding stellar mass. In the same panel we show two curves (red and blue dashed lines) representing a compilation of data from Rodrigues-Puebla et al. (submitted), after correcting for the different adopted IMF. Their molecular Hydrogen masses are derived using CO as a main proxy, taking in account upper limit cases when CO is not detected. This updates the curves over-plotted in Figure 8 from S18 (and also those published by \citealt{Calette.etal.2018}) after a change in the calculated mass functions. The same separation from RGs to SFGs is appreciated when dividing galaxies for morphological types (early- and late-type), with late-type galaxies being significantly gas richer than early-type, in particular at high stellar masses. Our results agree with those ones, showing that RG do not form stars (or form at a really low pace) due to their lack of gas in comparison with SFGs of the same stellar mass. However, as indicated before, there must be another mechanism
on top of the lack of gas, since RGs and GVGs have lower SFE. Thus, for a given M$_{\rm gas}$ they have lower SFRs too.

Only one AGN host is found above our best fit of the distribution, IC 1481. This galaxy, as mentioned before (c.f. \cref{sec:analysis:candidates:sample}), is a candidate to be hosting an outflow by the work of \citealt{LopezCoba.etal.2019}, with 43 per cent of its extra-planar spaxels ionized by shocks. However, in the same study, they indicate that 56 per cent of the explored area is probably ionized by an AGN. The qualitative results regarding the AGN hosts are the same reported by S18 for a MaNGA sample. As we see in the \ulpan\ of \cref{fig:prop_morph}, the AGN hosts are at the high-mass end of the stellar-mass distribution. Besides the spread, they are also mostly at the high-mass end of the molecular gas mass distribution too, but usually with less gas than the SFGs at same integrated stellar-mass. One more time suggesting that even having plenty of molecular gas the AGN appears to be acting as a SF strangler of their host galaxies.

\section{Discussion}
\label{sec:disc}

\begin{figure*}
  \includegraphics[width=\columnwidth]{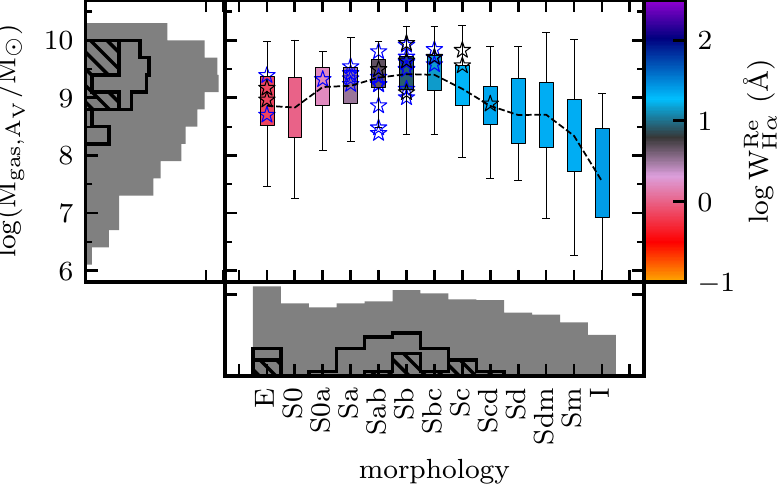}
  \includegraphics[width=\columnwidth]{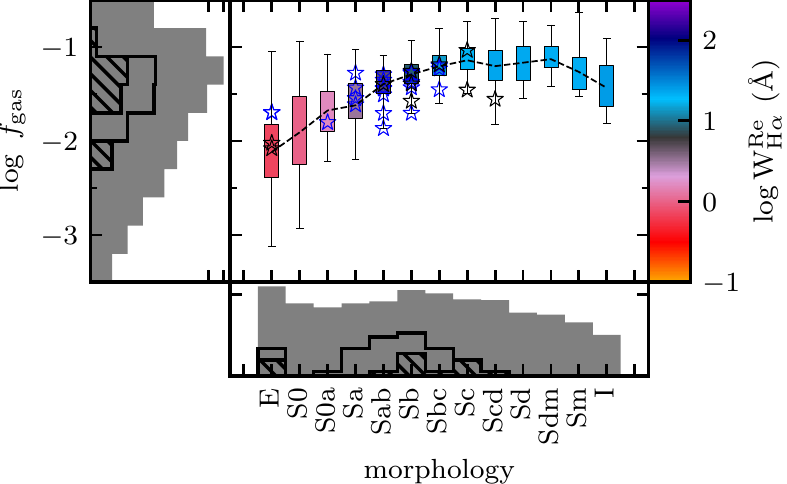}
  \includegraphics[width=\columnwidth]{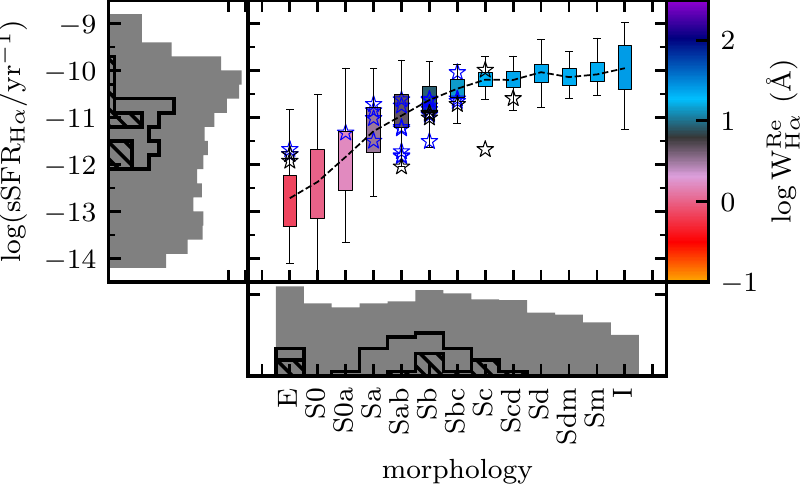}
  \includegraphics[width=\columnwidth]{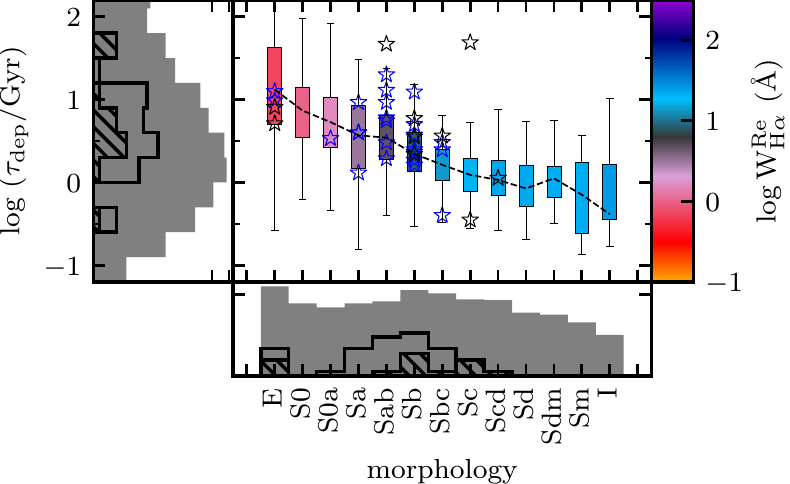}
  \caption{Box plots from integrated molecular gas mass, gas fraction, sSFR and molecular gas depletion time distributed in different morphological types, as \cref{fig:prop_morph}.
    \ULpan: Integrated molecular gas mass derived from dust extinction following S18, ${\rm M}_{\rm gas,A_V}$.
    \URpan: The integrated molecular gas fraction, $f_{\rm gas}={\rm M}_{\rm gas,A_V}/({\rm M}_{\rm gas,A_V} + {\rm M}_\star)$.
    \LLpan: The integrated sSFR (as \rpan\ in \cref{fig:SFMS}).
    \LRpan: The molecular gas depletion time, derived as the inverse of the star-formation efficiency, $\tau_{\rm dep}=1/{\rm SFE}={\rm M}_{\rm gas}/$SFR.
    It presents the same design (colors and symbols) as \cref{fig:prop_morph}. The color figure can be viewed online.}
  \label{fig:morph_gas}
\end{figure*}

\subsection{Comparison with other AGN classification schemes for IFS data}
\label{sec:disc:compare}

This article uses a similar scheme of AGN hosts identification adopted in S18 for the MPL-5 sample. Comparisons with S18 results are distributed among this entire article. However, recent works did the same job of classifying AGN hosts and it is worth to notice the main differences between the schemes and resulting selections.

Using a similar MPL-5 sample, \citet{Wylezalek.etal.2018} identified 303 AGN hosts, a number three time bigger than that of S18. They classify the hosts based in a spaxel-by-spaxel analysis of [NII]/\Ha vs [OIII]/\Hb and [SII]/\Ha vs [OIII]/\Hb, including the equivalent width of \Ha. In addition, they use the distance of those spaxels classified as Seyfert/LINER and the K06 curve in the [SII]/\Ha vs [OIII]/\Hb plane. 
This type of scrutiny considering all spaxels from a galaxy adds up to the final number of possible AGN hosts some objects that suffer from ionization by outflows and also from recently dimmed (or 'just' turned off) AGNs. Furthermore, the MaNGA sample includes a color-enhanced subsample in order to balance the colour distribution at a fixed stellar mass. However, this selection also populates
the GV with spiral galaxies with a high number of inclined galaxies enhancing the number of detected outflows. This could be visualized in the archetypal galaxies shown in their article, (e.g, Figures 1, 4 and 7 of \citealt{Wylezalek.etal.2018}). The example in their Figure 1 is a clear example of this caveat. Analyzing this galaxy with the same process done by \citet{Wylezalek.etal.2018} we also would classify this galaxy as a probable AGN host, but when looking towards the galaxy center, we see a region ionized by star-formation with a textbook biconical outflow that goes towards the outer parts. This is confirmed by the enhance of the ratio between velocity-to-velocity dispersion (v/$\sigma$) parameter in the center and the outflow region. Moreover, when adding the statistics of identifyied outflows by \citet{LopezCoba.etal.2019} and removing the overlapped galaxies, the number of identified candidates increases up to $\sim$6 per cent of galaxies accounting for shock ionization. Therefore, this effect helps to account for the high number of galaxies identified as harbouring an AGN in this study. 

Four recent studies using the same MPL-5 sample identify 62 AGN hosts using the SDSS-III DR12 \citep{Alam.etal.2015} integrated nuclear spectra, that comprises the central $3'' \times 3''$ region of the select galaxies. \citet{Rembold.etal.2017} presents the sample and a study of the nuclear stellar populations with \starlight\ \citep{CF.etal.2005}. Their classification is based on the BPT diagram togheter with the WHAN diagram, i.e., less restrictive regarding the hardness of the ionization field (they do not use \sii/\Ha and \oi/\Ha ratios in the classification scheme). Also, they apply the same lower-limit of \WHacen for AGN hosts used in this study (3\,\AA). The emission lines fluxes and equivalent widths measurements utilized in the classification were derived by \citet{Thomas.etal.2013}. Comparing the AGN sample with a control sample with two non-active galaxies for each active one, they found that in the luminosity of \oiii, L(\oiii), is enhanced in strong luminous AGNs, also identifying an increasing fraction of young stellar populations in the centre of those galaxies hosting strong AGNs and also a decreasing fraction of the older ones. This kind of balance between the old-to-young stellar populations suggests that most luminous AGNs were trigged by a recent supply of molecular gas that produces SF in the circumvent area of the AGN. This support a connection between bulge growth via formation of new stars and the growth of the SMBH via accretion during the AGN phase of activity. For the kinematics, \citet{Ilha.etal.2019} find an enhancement in the differences between the velocity dispersion of the gas and the stars for AGN hosts in the central kpc. The cause of this displacement is understand as the gas close to the galaxy centre is being perturbed by outflows ejected from an active nucleus. Indeed some of our AGN hosts present detectable outflows, with one of them also reported by \citet{LopezCoba.etal.2019} as a host of an AGN-driven outflow. Other 

\citet{Mallmann.etal.2018} studied the resolved properties of the stellar populations and \citet{doNascimento.etal.2019}, the gas excitation and the SFR distribution. Although, the derivation of the gas masses is different, they find that the AGN hosts have higher ionized gas masses, a result which is also reported in S18 and here in \cref{sec:results:prop:gas}. In addition, they said that many of the nuclear regions of MPL-5 galaxies would be classified as AGN in a BPT-only scheme, this reaffirms the importance of the inclusion of the \WHa in the classification process. The SFR is derived from \Ha which includes ionization other than star-formation, reason why the SFR is not evaluated at the central regions. The average population of galaxy bulges is old and as we said in many parts of this study, the \Ha emission is contamined by diffuse ionization by HOLMES. When an active nucleus is present, the hard AGN ionizing spectrum dominates the \Ha emission, however in very dusty AGNs this effect could be compensated. They observe higher differences in the total SFR between AGNs and the control sample within early-type galaxies. The reason could be that the gas that ignites AGNs in early-type galaxies also ignites some star-formation in their hosts disks. For late-type hosts they also do not see evidence for a SF-quench caused by the AGN. However, both gas and stellar population studies are complementary and support an inside-out quench of the SF in the average population of galaxies. Otherwise, in AGN hosts an outside-in scenario could describe the recent SF. We find no significant difference in the integrated SFR between Type-I and Type-II AGN hosts. This could be explained if both are the same type of AGNs only seen by different angles. In the other hand, the integrated spectra is dominated by the light of the young stars which live in the disks dictating the value of the integrated SFR. Although this study only includes central, effective radii and integrated properties, we find that the difference between \WHacen and \WHaRe is bigger for AGN hosts ($\approx$10\AA) than non-active galaxies ($\approx$2\AA), with this behaviour persisting when dividing by ionization classes (SFG, GVG and RG): for all SFGs the $\Delta(\WHa)$ is 3.08\AA\ and 21.87\AA\ for SFG AGN hosts; for GVGs, 4.25\AA\ and 7.44\AA\ for GVG AGN hosts; and for RGs, 0.91\AA\ and 4.07\AA\ for RGs AGN hosts.

\begin{table*}
    \centering
    \caption{Distribution of mean values of properties from \cref{fig:morph_gas} along morphological classes.}
    \begin{tabular}{c|c|c|c|c}
        \hline
        Morph. type & $\log({\rm M}_{\rm gas,A_V}$/M$_\odot)$ & $\log f_{\rm gas}$ & $\log$(sSFR$_{\Ha}/{\rm yr})$ & $\log(\tau_{\rm dep}$/Gyr) \\
        \hline
        \hline
        E   & 8.86 & -2.12 & -12.72 &  1.12 \\
        S0  & 8.83 & -1.91 & -12.37 &  0.87 \\
        S0a & 9.19 & -1.68 & -11.84 &  0.73 \\
        Sa  & 9.21 & -1.62 & -11.29 &  0.57 \\
        Sab & 9.35 & -1.39 & -10.96 &  0.54 \\
        Sb  & 9.41 & -1.29 & -10.63 &  0.34 \\
        Sbc & 9.40 & -1.20 & -10.39 &  0.21 \\
        Sc  & 9.16 & -1.14 & -10.20 &  0.09 \\
        Scd & 8.86 & -1.20 & -10.20 &  0.03 \\
        \hline
        Sd  & 8.70 & -1.17 & -10.04 & -0.07 \\
        Sdm & 8.71 & -1.13 & -10.14 &  0.05 \\
        Sm  & 8.34 & -1.26 & -10.08 & -0.15 \\
        I   & 7.55 & -1.44 &  -9.95 & -0.38 \\
        \hline
    \end{tabular}
    \label{tab:morph_gas}
\end{table*}

\subsection{The AGN and their hosts}
\label{sec:disc:AGNsCALIFA}

The results shown in this study indicate that AGN hosts are mostly located in the transition region between SFGs and RGs, broadly known as the GV. This occurs with basically all the integrated properties explored along this article. In this regards, we confirm previous similar explorations using different samples and observational techniques \citep[e.g.][S18]{Schawinski.etal.2014}. Two possible scenarios have been proposed to explain these results: (i) either the AGN halts the SF due to a removal or heating of the molecular gas, moving galaxies from the SF cloud to the RGs one; or (ii) an already retired galaxy suffers a rejuvenation by the inflow of gas that also feed the AGN \citep[e.g.][]{King.2005, Gaibler.etal.2012, Rovilos.etal.2012}. In both cases the AGN hosts would reside in the transition regime between both populations.

AGN hosts are among the most massive galaxies and those with denser central regions at each morphological class (c.f. \llpan\ of \cref{fig:prop_morph}). At the same time, compared with SFG, they are the ones with lower SFR and sSFR. Their mean stellar ages indicate that they are usually dominated by old stellar populations. When comparing their ages with those of SFGs and RGs, it seems that the AGN phase last $\sim$1-3 Gyr. In general it seems that the removal of molecular gas is the main effect of the presence of an AGN (either physical removal or heated preventing to form molecular gas). This is more clearly seen by a morphological segregation explored in \cref{fig:morph_gas} and summarized in \cref{tab:morph_gas}. In this figure we present the distribution of the integrated M$_{\rm gas}$, gas fraction ($f_{\rm gas}$ = ${\rm M}_{\rm gas}/({\rm M}_{\rm gas} + {\rm M}_\star)$), sSFR, and $\tau_{\rm dep}$ along the morphological type of galaxies. Exploring the properties of galaxies from late-to-early types, AGN hosts are mostly found in the peak of the estimated mean gas mass, gas fraction, and sSFR. Conversely, the molecular gas depletion time shows a smooth decline from early-to-late types, with AGN hosts following the same average sequence. From that, AGN activity and bulge growth appear to work concomitantly since the molecular gas fraction starts to decrease when AGN ignites and $\tau_{\rm dep}$ increasing with morphology \citep[e.g.][]{Cheung.etal.2012, Mancini.etal.2015}. Although most late-type galaxies have less gas mass, when normalizing it by the sum of stellar and molecular gas masses, their gas fraction is high and mostly constant till the Sbc galaxies, making those galaxies the ones with high sSFR and highest efficiency to form stars. The early-types are scant of gas in absolute and relative values, with almost no SF and with high $\tau_{\rm dep}$ (low SFE).

As mentioned in \cref{sec:results:prop:SFMS}, we inspect the offset in SFR, $\Delta$($\log$SFR), with respect to the SFMS for each galaxy along its stellar mass \citep[following,][]{Elbaz.etal.2007, Bitsakis.etal.2019} in order to infer how AGNs affects the SF in galaxies. Regarding this parameter, the scatter for the SFGs (that is almost zero by construction) appears to be independent of the gas fraction. However, for GVGs and RGs the gas fraction seems to have an influence in $\Delta$($\log$SFR), affecting more those galaxies in the transition from GVG to RG. At the same gas fraction AGN hosts are the more distant from the SFMS, with not much difference between Type-I and II. As was foreshadowed by the \lrpan\ of \cref{fig:morph_gas}, the influence of the AGN lowering the SFE appears to not be the main driver of the quenching. On the other hand, the AGN presence helps to heat/expel the molecular gas (c.f. \ulpan\ of \cref{fig:morph_gas}), or at least AGN hosts clearly show a deficit of this gas. The conclusions here point in the same direction suggested by S18, where a mixed scenario of quenching exists: (i) an AGN feedback quenching, where the AGN injects energy heating the cold gas and also decreasing its reservoir; and (ii) a morphological quenching, where the growth of the bulge helps to increase the depletion time of the molecular gas.

\section{Summary}
\label{sec:summary}

We explore the presence of AGNs in a sample of 867 galaxies in the nearby universe (z$\sim$0.04) extracted from the CALIFA survey. In order to characterize which galaxies host an AGN in comparison with the non active ones, we analyse the IFS data using {\sc Pipe3D} to derive the properties of the
underlying stellar population and the ionized gas. From the derived dataproducts we found the following results:

\begin{enumerate}
    \item We classify our galaxies regarding their SF stage using the equivalent width of \Ha at effective radius (\WHaRe), comprising 382 SFGs, 185 GVGs and 300 RGs.
    \item Using the information provided by the emission line ratios and the \WHa in the central region of the analysed galaxies we found 34 AGN host candidates, 10 of them classified as Type-I and 24 Type-II. They are distributed in morphology from E to Sc galaxies with a predilection to Sab+Sb galaxies. Type-I hosts are more concentrated in late-type galaxies apart from 2 found in elliptical ones, while Type-II are distributed along all morphological classes.
    \item Regarding the stellar properties, the AGN hosts are in the range of the most massive and centrally peaked (in stellar mass density) ones with respect to their non-active counterparts. They were found in GV-like locations in all analysed diagrams that present a bimodality in their values.
    \item Considering that morphology could represent a pathway for the history of galaxy evolution (from late to early-types), AGN hosts start to appear in morphological classes that present the higher molecular gas masses, and decreases with the smooth turn-down observed toward early-types. This change of behaviour could also be seen in the molecular gas fraction and the sSFR, but not in the depletion time (or SFE). Therefore, the AGN appear to have an important role in decreasing the molecular gas-fraction of their hosts but not in lowering the SFE (that seems to be more clearly related with morphology).
\end{enumerate}

All these results indicate that AGN observed in galaxies at the transition phase between SFGs and RGs. If there is a causal connection, it is still not clear. However, our results points towards in a mixed scenario where AGN feedback and growth of the bulge act toghether in order to first quench (by removing or heating the molecular gas) and then prevent (by decreasing the SFE) their hosts SF.


\section{Acknowledgements}
We thanks CONACYT FC-2016-01-1916 and CB-285080 projects and PAPIIT IN100519 project for support on this study. L.G. was funded by the European Union's Horizon 2020 research and innovation programme under the Marie Sk\l{}odowska-Curie grant agreement No. 839090. EADL thanks all Stellar Population Synthesis and Chemical Evolution group of the Instituto de Astronomía $-$ UNAM for the help through the by-eye morphological classification.


\bibliographystyle{mnras}
\bibliography{bibliography}

\appendix
\section{The integrated BPTVO}
\label{app:intBPTVO}
\begin{figure*}
    \includegraphics[width=1.0\textwidth]{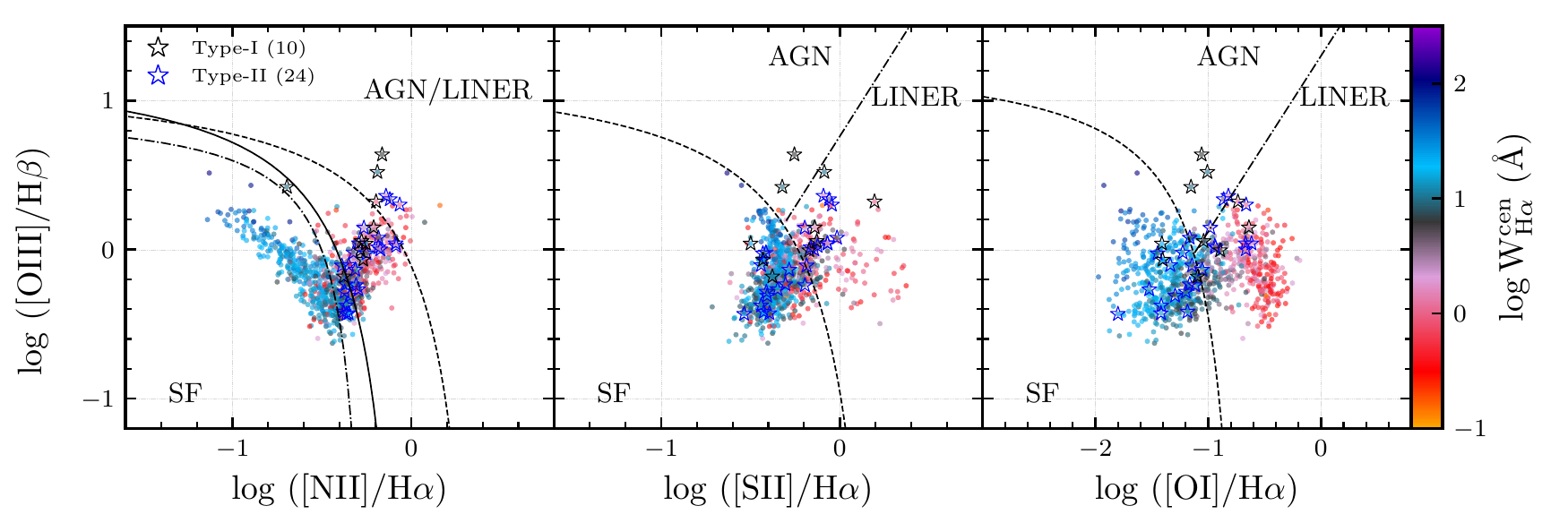}
    \caption{Diagnostic diagrams using optical emission lines ratios (\nii/\Ha, \sii/\Ha and \oi/\Ha versus \oiii/\Hb). Differently from \cref{fig:BPTVO}, the emission lines are measured at each galaxy integrated spectrum (i.e. entire field-of-view). In addition, if the integrated data were used to classify the AGN hosts, we would have been classified 4 AGNs, 3 Type-I and 1 Type-II. The markers, lines and colors are the same as \cref{fig:BPTVO}. The color figure can be viewed online.}
    \label{fig:intBPTVO}
\end{figure*}

In order to classify AGN hosts, instead of using the spectra integrated in the central $3''\times3''$ area, we could use the emission line ratios from the integrated data cube. We show this experiment in \cref{fig:intBPTVO}, the same as \cref{fig:BPTVO} but considering the ratios from the integrated spectra. It is clear that the AGN influence does not dominate the entire FoV of galaxies. Regarding IFS from fields that encase entire galaxies, the integrated spectrum is dominated by the light of the stellar components, this could be seen because almost all Type-II AGN hosts have their line ratios dominated by SF, as one could see both \sii/\Ha and \oi/\Ha vs \oiii/\Hb diagrams. If the method of AGN hosts classification employed in this article were feed by integrated data we would have been classified only 4 AGN hosts, 3 Type-I and 1 Type-II, a number one order lower than the actual number we report. 

\bsp	
\label{lastpage}
\end{document}